\documentclass[%
 reprint,
superscriptaddress,
 amsmath,
 amssymb,
 aps,
 prd,
]{revtex4-2}

\usepackage{graphicx}
\usepackage{dcolumn}
\usepackage{bm}
\usepackage{color}
\usepackage{amsthm} 
\usepackage{wasysym}
\usepackage{mathtools}

\usepackage{amsmath} 
\usepackage{amsthm} 
\usepackage{amssymb}	
\usepackage{graphicx} 

\usepackage{bm}
\usepackage{tikz}
\usepackage{color}
\usepackage{wasysym}
\usepackage{mathtools}
\usepackage{simplewick}

\usepackage{enumitem}
\usepackage[percent]{overpic}
\usepackage[caption=false]{subfig}
\usepackage{rotating}
\usepackage{hyperref}
\usepackage{cancel}
\usepackage{comment}

\allowdisplaybreaks

\newcommand{\dd}{\mathrm{d}}
\newcommand{\Tr}[0]{\text{Tr}}
\newcommand{\ii}{\mathrm{i}}
\newcommand{\abs}[1]{\left| #1 \right|} 

\newcommand{\ket}[1]{\left| #1 \right>} 
\newcommand{\bra}[1]{\left< #1 \right|} 
 \newcommand{\ketbra}[2]{\left| #1 \vphantom{#2}\right>\!\!\left< #2\vphantom{#1}\right|}

\begin{document}


\title{Locality and entanglement harvesting in covariantly bandlimited scalar fields}

\author{Nicholas Funai}
\affiliation{Centre for Quantum Computation and Communication Technology, School of Science,
RMIT University, Melbourne, Victoria 3001, Australia}%


\author{Nicolas C. Menicucci}
\affiliation{Centre for Quantum Computation and Communication Technology, School of Science,
RMIT University, Melbourne, Victoria 3001, Australia}%


\date{\today}

\begin{abstract}
Considerations of high energies in quantum field theories on smooth manifolds have led to generalized uncertainty principles and the possibility of a physical minimal length in quantum gravitational scenarios. In these models, the minimal length would be a physical limit, not just a mathematical tool, and should be Lorentz invariant. In this paper, we study two-qubit communication and entanglement harvesting in a field subject to a covariant bandlimit (minimum length) and present the changes induced by this bandlimit. We find the bandlimit introduces nonlocality and acausal communication in a manner unlike non-covariant bandlimits or other quantum optical approximations. We also observe that this covariant bandlimit introduces uncertainties in time and temporal ordering with the unusual behavior attributed to the behavior of virtual particles being modified by the covariant cutoff.
\end{abstract}

\maketitle



\section{Introduction}

The long-standing open problem of physics over the century has been the unification of quantum physics and general relativity under a single consistent unifying theory. The extremely weak coupling of gravity has long been an obstacle to empirical measurements for a unifying theory, requiring very high particle energies or very strong gravitational fields, both of which are currently unavailable on Earth. However, it is expected that some breakdown in the definiteness of space and time will occur in this currently inaccessible region as a result of fluctuations in spacetime curvature. This idea has been molded into a thought experiment from which generalized uncertainty principles~(GUPs) are derived, i.e.,~modified Heisenberg uncertainty relations that include the effects of spacetime curvature~\cite{Hossenfelder2013}. These GUPs then imply the existence of a minimum length scale~\cite{Pye_2019}.

The introduction of minimum lengths is not new in field theory, with ultraviolet~(UV) cutoffs often being used for quantum field theory~(QFT) renormalization or in condensed-matter lattice theories. In the former case, the UV cutoff is used as a mathematical tool to circumvent divergences rather than as a physical limit. The latter case uses the physical cutoff associated with the crystal's lattice spacing.

Recent works have begun considering the effects of a physical minimum length on QFT interactions and protocols, especially searching for low-energy signatures that might refine current speculations of high-energy field behaviors. Studies by Mart\'{i}n-Mart\'{i}nez and colleagues considered the effects of UV cutoffs in relativistic quantum information~(RQI) protocols, demonstrating a loss of causality in such situations, as well as the change in an Unruh-DeWitt~(UDW) detector's response in the presence of a UV cutoff~\cite{PhysRevD.92.104019}. Belenchia
et al.~\cite{PhysRevD.94.061902} found that under certain conditions, these detectors could detect a UV cutoff six orders of magnitude greater than its dominant energy scale.

The recent application of Shannon sampling theory~\cite{6773024} to UV cutoff QFTs~\cite{PhysRevLett.85.2873} has led to a resurgence of interest in minimum length models and their broader implications. This information theoretic approach to bandlimited QFT has been explored by Pye et al.~\cite{PhysRevD.92.105022} and Henderson et al.~\cite{PhysRevD.102.125026} who explore the changes in locality and field entanglement introduced by the cutoff. Henderson and Menicucci's~\cite{PhysRevD.102.125026} work explores these issues via entanglement harvesting, finding an increase in entanglement harvesting range, consistent with the previously studied notions of causality violation. Through Shannon sampling Henderson and Menicucci were also able to reinterpret the (UV cutoff induced) causality violation instead as a detector nonlocality in a non-cutoff field. This suggested that a QFT minimum length would simply blur the locality of detectors, with almost no appreciable changes for macroscopic observers. 

The work of Lewis et al.~\cite{PhysRevD.108.096024} also explored non-covariantly bandlimited QFTs, focusing on the continuous symmetries of the Shannon sampling lattice. Despite the lattice description of a bandlimited QFT, Lewis et al. showed that continuous symmetries are inherited from the continuous representation, as are their respective Noether conserved quantities. This combination of Shannon sampling and QFT has helped bridge the chasm between the inherent discreteness of spacetime normally assumed (but then renormalized away) in ordinary QFT and the smoothness of spacetime manifolds assumed in general relativity~(GR).

However, a drawback of past works, including those mentioned above, is the lack of Lorentz symmetry to the choice of UV cutoffs applied. A succession of publications~\cite{PhysRevLett.92.221301, 10.1063/1.4790482, PhysRevLett.119.031301} put forth a means of introducing a Lorentz-covariant cutoff on a QFT. This covariant cutoff can be derived from GUPs~\cite{Pye_2019} and has non-trivial effects, least of which is the absence of a (3+1)-dimensional Shannon sampling lattice~\cite{Pye_2023}. This cutoff acts trivially on the field operators themselves (e.g.,~$\hat{\phi}$) but instead effectively modifies the definition of time ordering that is used to define the Feynman propagator~\cite{Pye_thesis}.

In other words, we have an important contrast: (1)~Non-covariant bandlimited QFTs merely change the state space of the field (and thus the field operator~$\hat{\phi}$ but leave the laws of quantum physics otherwise unchanged. (2)~Covariant bandlimited QFTs are a modification to the laws of quantum time evolution (by modifying the definition of the Dyson series). Effects of the latter appear when considering the scattering amplitudes of Feynman diagrams that involve contributions from virtual-particle (off-shell) excitations.


This manuscript will study the effects of a covariant UV cutoff on simple two-detector RQI protocols, namely communication and entanglement harvesting. We choose two-detector RQI protocols as they are one of the simplest (2nd-order) perturbative models where a measurable signature of the covariant bandlimitation may be observed.


The communication protocol~\cite{PhysRevD.92.104019} is simple and straightforward. It is an unambiguous tool for determining if a system is local and causal and for crudely quantifying the violations when it is not. Entanglement harvesting~\cite{VALENTINI1991321, Reznik2003, Martin-Martinez_2014,PhysRevD.94.064074} provides more descriptive quantitative information, making it a popular and useful tool for probing QFTs with RQI tools. This, in conjunction with the communication protocol, can provide a detailed description of the quantum field and the changes induced by the introduction of a covariant UV cutoff, whilst providing some useful intuition and physical interpretations for these effects.

In section~\ref{sec_z2}, we introduce a covariant UV cutoff to a QFT and discuss superficial changes induced by the cutoff. In section~\ref{sec_z3}, we briefly overview the two-qubit communication protocol and then explore the changes in locality and causality introduced by the UV cutoff. In section~\ref{sec_z4}, we consider the effects of the covariant UV cutoff on the entanglement harvesting protocol with both qualitative and some quantitative analysis. Finally, in section~\ref{sec_z5}, we discuss the results and their broader physical consequences.

Note that throughout this manuscript we will be using the terms UV cutoff and bandlimitation interchangeably. Also, we are only considering massless quantum fields.

\section{Covariant UV cutoffs in QFT}\label{sec_z2}
\subsection{Preliminary: Propagators (Green's functions) in QFT}
Due to the important role of propagators in covariant bandlimitation, we will briefly review QFT from a propagator perspective.

As we know, Green's functions are useful tools for solving inhomogeneous differential equations where the inhomogeneity usually describes some physical source in an otherwise (usually) conservative system~\cite{haberman2018applied}. Green's functions are solutions for pointlike sources (Dirac deltas) and are often called propagators in the context of wave equations. 

In the case of a (3+1~D) scalar field theory, the propagators take the general form 
\begin{align}
G(x,x')&=\int\limits_{\mathcal{C}}\frac{\dd^{4}k}{(2\pi)^{4}}\frac{e^{-\ii k_{\mu}(x-x')^{\mu}}}{k_{\mu}k^{\mu}-m^{2}}
\end{align}
where $\mathcal{C}=\mathbb{R}^{3+1}$ with different $k_{0}$~contours around the singularities corresponding to different boundary conditions~\cite{Birrell1984}. While the propagator's \textit{raison d'\^{e}tre} is to model classical source perturbations in quantum fields, its functionality persists in interacting theories, e.g. $\phi^{4}$ or UDW-interactions~\cite{UDW_1,UDW_2}. One such use involves a truncated (perturbative) use of Picard's method of successive approximations~\cite{picard1926traite} that yields an equivalent result to the usual canonical perturbation theory while avoiding the additional operator and state structure used in canonical quantization. 

The direct Green's function approach to solving the Klein-Gordon equation is uncommon, but it is implicitly used in both canonical and path-integral quantization through Wick's theorem. In RQI protocols, 2nd-order perturbation theory is commonly used and the Green's functions are implicitly present in the time-ordered 2-point correlators. Propagators and Green's functions are omnipresent in interacting QFTs, even if their form is unconventional. This is important to remember as the simplest implementation of the covariant cutoff is via a direct modification of the form of the propagators, as shown in the following section.

It should also be noted that the covariant cutoff's modification of the Green's function will also change the right-hand side of the Green's function differential equation. The usual pointlike sources (Dirac deltas) will be modified and become some decaying oscillatory function in spacetime~\eqref{eqaa6}, consistent with a covariant minimum length scale.

\subsection{Implementing a covariant cutoff}\label{sec2b}
Non-covariant UV cutoffs are commonly used in interacting QFTs~\cite{peskin} and condensed-matter systems~\cite{altland2010condensed}, where restrictions are applied to the integral domain of the field's mode decomposition. These cutoffs suffer from a lack of Lorentz covariance, which is particularly inconvenient when dealing with relativistic quantum fields in relativistic sensitive settings, e.g.,~Unruh effect~\cite{Unruh_effect_1,Unruh_effect_2}. Chatwin-Davies, Kempf and Martin~\cite{PhysRevLett.92.221301, 10.1063/1.4790482, PhysRevLett.119.031301} introduced the notion of covariant UV cutoffs in the path-integral quantization formalism, and later Pye translated covariant UV cutoffs into the canonical quantization formalism~\cite{Pye_thesis}. We will work in the canonical quantization formalism. In this section, we will overview the covariant UV cutoff (see Appendix~\ref{seca_appendix1} or Pye's thesis~\cite{Pye_thesis} for further details).

\begin{figure}[!t]
\includegraphics[width=0.9\columnwidth]{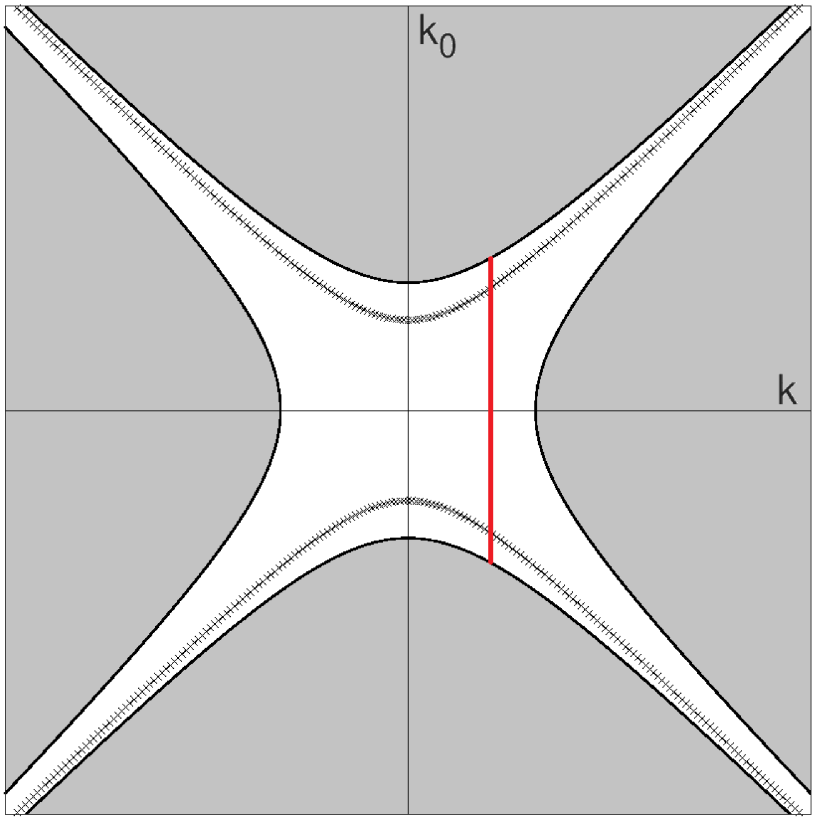}
\caption{\label{fig0_UV}3+1 D integration domain after covariant UV cutoff limitations. $k_{0}$ on the y-axis and $\bm{k}$ on the x-axis. The hyperbolic lines represent specific mass shells. The vertical line is the new integration domain of $k_{0}$ under the covariant UV cutoff. Note that for large enough $\bm{k}$ the integration domain becomes disjointed. Figure from \cite{Pye_thesis}.}
\end{figure}

At the core of the covariant UV cutoff is the eigenvalue equation
\begin{align}
\square f_{\lambda}&=\lambda f_{\lambda},\label{eq1}
\end{align}
where $\square$ is the d'Alembertian operator, $f_{\lambda}\in \text{L}_{2}(\mathbb{R}^{3+1})$ and $\lambda\in\mathbb{R}$. The covariant UV cutoff involves restricting the working space of functions to those spanned by eigenfunctions whose eigenvalues satisfy $\abs{\lambda}<\Lambda^{2}$, where $\lambda=\abs{\bm{k}}^{2}-k_{0}^{2}$ (in flat spacetime). In flat spacetime, this cutoff can be visualized in figure~\ref{fig0_UV}. In path-integral quantization, this means restricting the functional integral's domain: $\Pi_{\Lambda} (\text{L}_{2}(\mathbb{R}^{3+1}))=\text{span}\{f_{\lambda}:\abs{\lambda}<\Lambda^{2}\}$, where $\Pi_{\Lambda}$ is a projector onto the space of covariantly bandlimited functions. In canonical quantization, this UV cutoff blurs the action of time ordering in the Dyson series. This mathematically manifests as modifications to the Feynman propagator and the retarded and advanced Green's functions, e.g.,~$G_{\textsc{f}}(x,x')\rightarrow \Pi_{\Lambda}G_{\textsc{f}}(x,x')\eqqcolon G_{\textsc{f},\Lambda}$,
\begin{align}
G_{\textsc{f}}(x,x')&=\int\limits_{\mathbb{R}^{3+1}}\frac{\dd^{4}k}{(2\pi)^{4}}\frac{e^{-\ii k_{\mu}(x-x')^{\mu}}}{k_{\mu}k^{\mu}-m^{2}+\ii\epsilon},\label{eq2}\\
G_{\textsc{f},\Lambda}(x,x')&=\int\limits_{\mathcal{B}}\frac{\dd^{4}k}{(2\pi)^{4}}\frac{e^{-\ii k_{\mu}(x-x')^{\mu}}}{k_{\mu}k^{\mu}-m^{2}+\ii\epsilon},
\end{align}
where $\mathcal{B}=\{k_{\mu}\in\mathbb{R}^{3+1}:\abs{k_{\mu}k^{\mu}}<\Lambda^{2}\}$ and $\epsilon\rightarrow 0^{+}$ indicates the path of the $k^{0}$ integration contour around the singularities. Mathematically, the cutoff prevents the $k^{0}$ integration contour from closing via $\pm \ii\infty$, disallowing the use of the residue theorem and requiring the calculation of the integral contributions from the non-singular integration path (i.e. principal value). These principal value contributions will be off-shell i.e. the propagator has non-zero Fourier coefficients $c(k_{\mu})$ when $k_{0}^{2}\neq \abs{\bm{k}}^{2}+m^{2}$. As mentioned in the previous section the propagators play an important role in evaluating observables, therefore these modifications will result in notable physical changes. Note that 2-point correlation functions without time ordering, e.g. Wightman function, are unaffected by this covariant UV cutoff.

In RQI protocols, it is uncommon to explicitly use Green's functions, especially since RQI protocols tend to be studied using 2nd-order perturbation theory~\cite{PhysRevD.105.065003,PhysRevD.109.045013}. To help form an intuition for how covariant UV cutoffs act we consider a 2nd-order Dyson expansion. In these cases, non-trivial UV cutoff effects are felt on the time-ordered terms only. Using the Fourier representation of the Heaviside step function,
\begin{align}
\Theta(t)&=\lim_{\epsilon\rightarrow 0^{+}}\frac{1}{2\pi\ii}\int\limits_{-\infty}^{\infty}\frac{\dd\nu}{\nu-\ii\epsilon}e^{\ii\nu t},
\end{align}
the covariant UV cutoff's effect on 2nd-order Dyson terms takes the form
\begin{align}
\Pi_{\Lambda} &[\Theta(s-s')\left<\vphantom{\phi}\right.\hspace{-0.6mm}\hat{\phi}(s,\bm{x})\hat{\phi}(s',\bm{x}')\left.\hspace{-0.6mm}\vphantom{\phi}\right>]
=\frac{1}{2\pi\ii}\int\frac{\dd\nu \dd^{3}\bm{k}}{(2\pi)^{3}2\omega}\label{eqa5}\\
&\times e^{\ii(\nu-\omega)(s-s')+\ii\bm{k}\cdot(\bm{x}-\bm{x}')}\Theta\left(\Lambda^{2}-\abs{(\nu-\omega)^{2}-\abs{\bm{k}}^{2}}\right),\nonumber\\
&=[(\Theta D^{+})*\delta_{\Lambda}](\Delta x),
\end{align}
where $D^{+}$ is the Wightman function (two-point $\hat{\phi}$~correlator) and $\delta_{\Lambda}$ is the covariantly bandlimited $3+1$~D Dirac delta function
\begin{align}
\delta_{\Lambda}(x)&=\int\limits_{\mathcal{B}}
    \frac{\dd^{4}k}{(2\pi)^{4}}
    e^{\ii k_{\mu}x^{\mu}}.
    \end{align}
The $*$ is a 3+1~D convolution, which helps visualize the effect of covariant bandlimitation on time-ordered correlators (and other general off-shell functions). See Appendix~\ref{seca_appendix1} for a more rigorous mathematical description. 

For higher order perturbative expansions this \hbox{$\delta_{\Lambda}$-convolution} method is ill-defined as it is unclear which set of perturbative integration variables should be used in the convolution. Instead, the use of Wick's theorem is indicated and a direct substitution of covariantly bandlimited Feynman propagators. Note that the energy gap of the detector plays no role in the modification of the Green's function or the time-ordering function. 

The covariant cutoff's change in \eqref{eqa5} has smoothed the Heaviside step function and thereby softened the strict time ordering of the Dyson expansion, introducing contributions from time-disordered terms. Unlike non-covariant UV cutoffs and other approximation models that modify the Hilbert space of the field or change the interaction Hamiltonian form~\cite{PhysRevD.102.125026}, the covariant UV cutoff has modified the notion of time-ordering itself. However, the deformation of time ordering is a logical inevitability of the covariant bandlimit, which introduces a minimum spacetime distance, including uncertainties in time.

Higher-order perturbative expansions will require Wick's theorem for an unambiguous implementation of the covariant UV cutoff. Currently, there is no procedure for implementing this covariant UV cutoff non-perturbatively, e.g.,~for harmonic oscillator detectors~\cite{Brown_2014} or Gaussian QM approach~\cite{PhysRevD.96.065008}, although this is an avenue of future research.

\subsection{Immediate consequences of a covariant cutoff}\label{sec2.a}
The covariant cutoff described above does not modify the form of the canonical field operators, e.g.,~$\hat{\phi}$ and $\hat{\pi}$. These operators are on-shell---i.e.,~given a mode decomposition, we have $k_{\mu}k^{\mu}=m^{2}$, and therefore $\Pi_{\Lambda}\hat{\phi}(x)=\hat{\phi}(x)$ provided $m<\Lambda$. This particular cutoff forbids the existence of massive fields with $m\geq \Lambda$, which is a reasonable restriction. 

It is also important to note that the covariant cutoff acts trivially on commutation relations. For example, space-like separated field operators still commute in a covariantly cutoff model (Appendix~\ref{sec_a3}). As we show in \S\ref{sec3a}, communication is possible despite the micro-causality~\cite{microcausality_AQFT} of the field operators and consequently micro-causality of field operators is not a reliable indicator of communication in a covariantly cutoff model. 

Another immediate observation of this covariant cutoff is that in 3+1~D we cannot construct a faithful lattice representation of the field. The trivial action of the cutoff on canonical field operators results in an unbounded Fourier volume and the corresponding lattice representation requires a divergent Nyquist-Beurlinger-Landau sampling density~\cite{Landau1967}. The field propagators are modified by the cutoff but still have a divergent hyperbolic volume (in 3+1~D)~\cite{Pye_2023} and consequently also require an infinite sampling density.

The changes to the Feynman propagator show that some non-trivial (observable) effects arise from introducing a covariant cutoff. This effect is best described in Feynman diagrams as a restriction of the 4-momentum $(k_{0},\bm{k})$ that virtual particles can have, i.e.,~restrictions of how off-shell these particles can be (illustrated in figure~\ref{fig0_UV}). We expect that a covariant cutoff would influence observations in interacting theories, e.g. self-interacting models~($\phi^{4}$)~\cite{Pye_thesis} or detector-field interacting models~(Unruh-DeWitt interactions~\cite{UDW_1,UDW_2}), where detectors serve as sources and sinks for virtual particles.

\section{Communication in QFTs with covariant cutoffs}\label{sec_z3}
\subsection{Communication protocol}\label{seca3a}

Quantum field theory is a quantum theory imbued with a causal structure constructed to model subluminal and luminal excitations. In non-cutoff QFT this causal structure manifests as micro-causality,~i.e. field operators commute when space-like separated~\cite{microcausality_AQFT,PhysRevA.81.012330}.  Some past works have studied the changes in micro-causality under different field approximations---e.g.,~non-covariant UV cutoffs and single-mode approximations~\cite{PhysRevD.92.104019} or the rotating-wave approximation~\cite{PhysRevD.100.065021}---as a means of demonstrating a breakdown in causality. However, the covariant cutoff described above acts trivially on the canonical field operators, leaving the commutation relations unchanged. Instead, the cutoff relaxes the strict notion of time ordering (c.f. Dyson series) by introducing some time uncertainty. It is therefore necessary to consider a more complete and physical model of communication to determine if a covariant cutoff introduces acausal behavior.

\begin{figure}[!t]
\includegraphics[width=0.9\columnwidth]{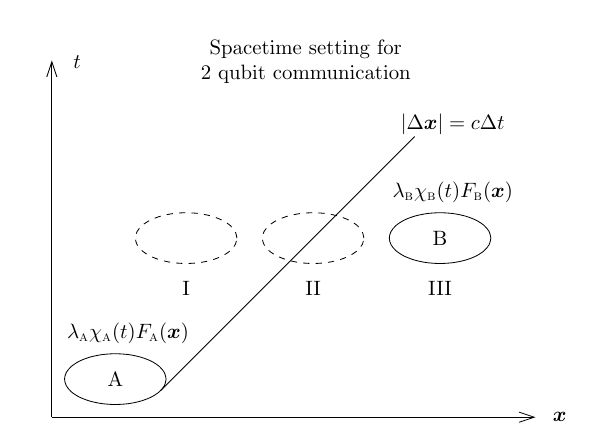}
\caption{\label{fig2_5} Setup for two-qubit communication. Alice's interaction has a compact spacetime support and Bob is placed at several different test locations to determine time-like, light-like, and space-like communication (regions I, II, III respectively). Note: the diagonal line $\abs{\Delta \bm{x}}=c\Delta t$ describes the boundary of causal (light speed) communication from Alice.}
\end{figure}

Consider the simple binary communication protocol between two UDW detectors and mediated by a massless QFT~\cite{PhysRevD.92.104019} (based on the Fermi problem~\cite{Fermi_prob}):
\begin{align}
\hat{H}_{\textsc{i}}&=\lambda_{\textsc{a}}\chi_{\textsc{a}}(t)\hat{\mu}_{\textsc{a}}(t)\otimes\int\dd^{3}\bm{x}\,F_{\textsc{a}}(\bm{x})\hat{\phi}(t,\bm{x})\nonumber\\
&\quad+\lambda_{\textsc{b}}\chi_{\textsc{b}}(t)\hat{\mu}_{\textsc{b}}(t)\otimes\int\dd^{3}\bm{x}\,F_{\textsc{b}}(\bm{x})\hat{\phi}(t,\bm{x}),\label{eq3}
\end{align}
where $\chi_{i}(t)$ are detector switching functions that modulate the strength of the interaction, $F_{i}(\bm{x})$ are the detector smearing functions that represent the size and location of the detectors, and $\hat{\mu}_{i}(t)$ is the monopole moment of the detector (usually $\hat{\sigma}_{x}(t)$ for qubit detectors with energy gap $\Omega$). The term `monopole moment' is used as Unruh-DeWitt detectors model light-matter interactions well when there is no exchange of angular momentum~\cite{PhysRevD.94.064074,PhysRevA.89.033835}. In this basic communication protocol, the sender (Alice) communicates a ``0'' by not interacting with the field ($\lambda_{\textsc{a}}=0$) and ``1'' by interacting with the field ($\lambda_{\textsc{a}}\neq 0$). 
The receiver (Bob) interacts with the field and must determine Alice's signal bit by studying his own reduced density matrix. In a causal theory, it is expected that Bob's reduced density matrix should be independent of Alice's signal ($\lambda_{\textsc{a}}$) if Alice and Bob are entirely space-like separated (i.e.,~no time-like or light-like paths exist between the respective spacetime supports of $\chi_{i}(t)F_{i}(\bm{x})$~\cite{sorkin1993,papageorgiou2023eliminating}). The relative placements of Alice and Bob under our consideration are illustrated in figure~\ref{fig2_5}. 

We consider the above communication protocol perturbatively, using the covariant cutoff action represented in \eqref{eqa5}.  Following the interactions of \eqref{eq3}, Bob's reduced density matrix will contain terms of the form (see Appendix~\ref{sec_b1} for more details)
\begin{align}
\hat{\rho}_{\textsc{b}}&=\Tr_{\textsc{a},\phi}(\hat{U}\hat{\rho}_{0}\hat{U}^{\dagger}),\\
\begin{split}
\hat{\rho}_{\textsc{b}}&=\hat{\rho}_{\textsc{b}}^{0}+\lambda_{\textsc{a}}\lambda_{\textsc{b}}\int\dd s\dd s' \dd^{3}\bm{x}\dd^{3}\bm{y}\,
F_{\textsc{a}}(\bm{x})F_{\textsc{b}}(\bm{y}) \chi_{\textsc{a}}(s)\chi_{\textsc{b}}(s')\\
&\quad\times\Theta(s-s')\left<\vphantom{\phi}\right.\hspace{-0.6mm}\hat{\phi}(s,\bm{x})\hat{\phi}(s',\bm{y})\left.\hspace{-0.6mm}\vphantom{\phi}\right>\mathcal{V}(s,\Omega_{\textsc{a}})
\hat{\sigma}_{x}^{\textsc{b}}(s')\hat{\rho}_{\textsc{b}}^{0}
+\dotsm,
\end{split}\label{eq5}
\end{align}
where $\mathcal{V}(s,\Omega_{\textsc{a}})= \alpha e^{\ii\Omega_{\textsc{a}}s}+\alpha^{*} e^{-\ii\Omega_{\textsc{a}}s}$ merely encodes the internal dynamics of Alice's qubit. In a causal theory, the $\lambda_{\textsc{a}}\lambda_{\textsc{b}}$-dependent terms sum to zero whenever Alice and Bob are space-like separated (region III in figure~\ref{fig2_5}). From \eqref{eqa5}, we can see the Heaviside step function and the 2-point field correlator are modified by introducing a covariant UV cutoff. This will result in acausal communication, even though space-like separated observables continue to commute, i.e. we see acausal communication despite the apparent maintenance of micro-causality.

\subsection{Acausal behavior}\label{sec3a}
To simplify the equations presented here we consider the initial state of the protocol to be pure and separable:
\begin{align}
\ket{\psi_{0}}=\ket{\psi_{\textsc{a}}}\otimes\ket{\psi_{\textsc{b}}}\otimes\ket{0},
\end{align}
with Alice and Bob's detectors in arbitrary pure states and the field initially in the vacuum and subject to the covariant cutoff ($\Lambda$). We also allow Bob to perform any projective measurement on his qubit after the communication protocol has taken place, and we consider the effect of Alice's signal bit on the probability distribution of Bob's measurement. Let $\ket{\psi}$ be one of the measurement basis elements. The probability of Bob (projectively) measuring $\ket{\psi}$ is given by (Appendix~\ref{sec_b1})
\begin{align}
\begin{split}
\bra{\psi}\hat{\rho}_{\textsc{b}}\ket{\psi}
&=\ii \lambda_{\textsc{a}}\lambda_{\textsc{b}} \int\dd s\dd s'\dd^{3}\bm{x}\dd^{3}\bm{y}\,F_{\textsc{a}}(\bm{x})F_{\textsc{b}}(\bm{y}) \chi_{\textsc{a}}(s)\chi_{\textsc{b}}(s')\\
&\quad\times \frac{\mathcal{W}(s,s')}{r}
\left\{\vphantom{\left[\int\limits_{L}^{r^{-}}+\int\limits_{r^{+}}^{R}\right]}
\underbrace{\delta(r-t)-\delta(r+t)}_{\text{Causal}} -\underbrace{\mathcal{I}_{\Lambda}(r,t)}_{\text{Acausal}}
\right\}\\
&\quad+\mathcal{P}(\lambda_{\textsc{b}}),
\end{split}\label{eq19}
\end{align}
where $\mathcal{W}$ encodes $\ket{\psi_{\textsc{a}}},\ket{\psi_{\textsc{b}}}$, $\ket{\psi}$, and the internal qubit dynamics; $\mathcal{P}$ is a 2nd-order polynomial in $\lambda_{\textsc{b}}$ only, $\hbox{t=s-s'}$, and $r=\abs{\bm{x}-\bm{y}}$. The $\delta$-functions correspond to causal (light-like) communication channels and $\mathcal{I}_{\Lambda}$ corresponds to the acausal communication channel. Note that $\mathcal{I}_{\Lambda}$ is the only term dependent on $\Lambda$ and can be expressed as
\begin{align}
\begin{split}
&\frac{\pi^{2}}{2 r}\mathcal{I}_{\Lambda}(r,t)=\\
&\int\limits_{\Lambda}^{\infty}\dd\omega\left(\frac{e^{-\ii\omega r}}{t^{2}-(r-\ii\epsilon)^{2}}+\frac{e^{\ii\omega r}}{t^{2}-(r+\ii\epsilon)^{2}}\right)
\frac{\cos( \sqrt{\omega^{2}-\Lambda^{2}}t)}{\sqrt{\omega^{2}-\Lambda^{2}}}\\
-&\int\limits_{0}^{\infty}\dd\omega\left(\frac{e^{\ii\omega r}}{t^{2}-(r+\ii\epsilon)^{2}}+\frac{e^{-\ii\omega r}}{t^{2}-(r-\ii\epsilon)^{2}}\right)\frac{\cos(\sqrt{\omega^{2}+\Lambda^{2}}t)}{\sqrt{\omega^{2}+\Lambda^{2}}}.
\end{split}\label{eq20}
\end{align}
In the limit $\Lambda\rightarrow \infty$ (i.e.,~non-cutoff field), $\abs{\mathcal{I}_{\Lambda}}\rightarrow 0$, which leaves \eqref{eq19} with the fully causal $\delta$-functions only. These $\delta$-functions are the mathematical descriptions of the strong Huygens' principle~\cite{McLenaghan1974,strong_huygen}, stating that a (flat) 3+1~D massless field only permits communication along light-like trajectories. Note that this general form of \eqref{eq19} tends to be found in QFTs with approximations, e.g.,~the rotating-wave approximation~\cite{PhysRevD.100.065021}, with the exact form of $\mathcal{I}_{\Lambda}$ dependent on the type of approximation or modification. It is generally expected that $\mathcal{I}_{\Lambda}$ should decay rapidly or decay and oscillate rapidly as $\abs{r-c\abs{t}}\rightarrow \infty$. This usually results in causality violations that occur on a small scale, e.g. minimum length ($\Lambda^{-1}$). 

\begin{figure}[!t]
\includegraphics[width=0.9\columnwidth]{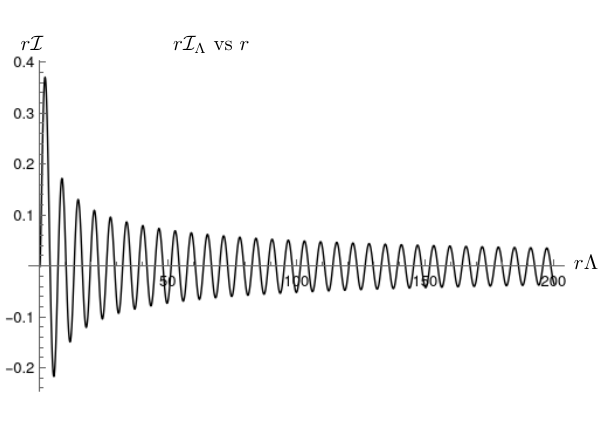}
\caption{\label{fig4} $r\mathcal{I}_{\Lambda}$ vs $r$ at $t=0$. $\mathcal{I}_{\Lambda}$ has a 1/r decay and has oscillatory behavior as with the non-covariant UV cutoff model. Note that the oscillation wavelength is on the order of the minimum length scale. Any atomic or even subatomic-sized detector, when convolved (Parseval–Plancherel identity) with $\mathcal{I}_{\Lambda}$, would only detect acausal behavior close to the lightcone (i.e. $r\approx 0$). We therefore expect any causality violations to be on the order of the minimum length scale. }
\end{figure}

The acausal term $\mathcal{I}_{\Lambda}$ is plotted in figure~\ref{fig4} ($r\mathcal{I}_{\Lambda}$ vs $r$) when $t=0$, where it is clear to see that $\mathcal{I}_{\Lambda}(r,t=0)$ introduces acausal signaling by means of some effective nonlocality. While the nonlocality of $\mathcal{I}_{\Lambda}$ is only polynomially decaying, its effects are currently difficult to observe since detector smearings and switchings are expected to be very large and smooth with very small Fourier coefficients near the dominant oscillation frequency of $\mathcal{I}_{\Lambda}$. From the Parseval–Plancherel identity, the overlap of the Fourier transformations of detector smearings, switchings and $\mathcal{I}_{\Lambda}$ will tend to be small, resulting in most nonlocal behavior integrating to virtually zero except when very close to the lightcone,  consistent with current experimental observations.

As mentioned above, we can interpret the covariant UV cutoff as truncating extreme off-shell virtual particle excitations (see figure~\ref{fig0_UV}). The non-trivial acausal term illustrated in figure~\ref{fig4} demonstrates that the extreme off-shell virtual particles play an important role in destructively interfering with near on-shell virtual particles thereby maintaining causality, a common interpretation in causality violating approximations. 

The fact that $\mathcal{I}_{\Lambda}$ is generated by virtual particles raises an interesting point regarding the statistics of Bob's detector. A na\"{i}ve evaluation of field expectation values, e.g. $\left<\vphantom{\phi}\right.\hspace{-0.6mm}\hat{\phi}\left.\hspace{-0.6mm}\vphantom{\phi}\right>$, will depend only on the real particle (i.e. causal) excitations generated by an interaction. This occurs since the observable is not subject to any time-ordering operator and is therefore unmodified by the covariant cutoff (Appendix~\ref{sec_a4}). Considering this fact with the dynamics of Bob's detector we see that, unlike non-cutoff fields~\cite{PhysRevD.103.125021}, complete knowledge of field expectation values is insufficient for modeling detector interactions in a covariant cutoff model. A Green's function perspective suggests that observables should experience some modification propagated from the initial detector interaction that might capture the behavior of acausal virtual particles. This would probably result in some dressed operators.

The result above shows that under a covariant UV cutoff, Alice and Bob can communicate despite being space-like separated, and curiously, the bare field operators over the same spacetime regions continue to commute (micro-causality). Quantitatively, acausal communication occurs near the lightcone boundary. This can be interpreted as the covariant cutoff softening the lightcone boundary on the minimum length scale.  Unlike other field approximations or modifications, the mathematical form of the detectors' reduced density matrix has changed, allowing for space-like communication despite the maintenance of micro-causality. This is achieved by introducing time uncertainty and modifying the time-ordering operator in the Dyson series. This distinguishes the covariant UV cutoff as an unusual modification to field theory and general quantum mechanics.

\section{Entanglement Harvesting}\label{sec_z4}
\subsection{Protocol}\label{sec4a}
Entanglement harvesting is a protocol first developed by Valentini in 1991~\cite{VALENTINI1991321} and then independently by Reznik in 2003~\cite{Reznik2003}, designed to probe the correlations and entanglement present in relativistic quantum fields while avoiding issues inherent with local projective measurements performed directly on relativistic quantum fields~\cite{sorkin1993,papageorgiou2023eliminating}. Entanglement harvesting is a staple protocol of relativistic quantum information used in several different ways~\cite{PhysRevD.94.064074,PhysRevD.96.065008,PhysRevD.98.085007,PhysRevD.96.085012}. As well as operationally demonstrating the underlying entanglement of the vacuum state, it has become a commonly used tool for probing the nonlocal attributes of QFTs, e.g.,~spacetime curvature~\cite{PhysRevD.79.044027,robbins2020entanglement,HENDERSON2020135732,hendersonEntanglingDetectorsAntide2019,Martin-Martinez_2014}. 

An entanglement harvesting protocol involves two detectors (qubits) A and B, with energy gaps $\Omega_{\textsc{a,b}}$, interacting with an initial vacuum field state via the UDW interaction \eqref{eq3}. We rewrite the interaction Hamiltonian specifically for qubit detectors:
\begin{align}
\hat{H}_{\textsc{i}}(t)&=\sum_{\mathclap{\textsc{d}=\{\textsc{a,b}\}}}
\lambda_{\textsc{d}}\chi_{\textsc{d}}(t)\hat{\sigma}_{x,\textsc{d}}(t)
\int\dd^{3}\bm{x}\,F_{\textsc{d}}(\bm{x})\hat{\phi}(t,\bm{x}).\label{eq7}
\end{align}
The expression above is in the interaction picture and assumes the detectors are stationary on a flat spacetime, which is a simple model that suffices to illustrate the covariant UV cutoff effects we wish to study. The post-interaction states are calculated via 2nd-order perturbation theory~\cite{peskin} after which the field is traced out, leaving a reduced density matrix:
\begin{align}
\hat{\rho}_{\textsc{ab}}&=
\begin{pmatrix}
1-\mathcal{L}_{\textsc{aa}}-\mathcal{L}_{\textsc{bb}} & 0 & 0 & \mathcal{M}^{*}\\
0 & \mathcal{L}_{\textsc{aa}} & \mathcal{L}_{\textsc{ab}} & 0 \\
0 & \mathcal{L}_{\textsc{ba}} & \mathcal{L}_{\textsc{bb}} & 0 \\
\mathcal{M} & 0 & 0 & 0
\end{pmatrix}
+\mathcal{O}(\lambda^{4}),
\end{align}
where (Appendix~\ref{sec_b2})
\begin{align} 
\mathcal{L}_{ij}&=\lambda_{i}\lambda_{j}\int\dd s\dd s'\dd^{3}\bm{x}\dd^{3}\bm{x}'\,e^{\ii\Omega_{i}s-\ii\Omega_{j}s'}\chi_{i}(s)\chi_{j}(s')\nonumber\\
&\quad\times F_{i}(\bm{x})F_{j}(\bm{x}')
\left<\vphantom{\phi}\right.\hspace{-0.6mm}\hat{\phi}(s',\bm{x}')\hat{\phi}(s,\bm{x})\left.\hspace{-0.6mm}\vphantom{\phi}\right>,\label{eqa17}\\
\mathcal{M}&=-\lambda_{\textsc{a}}\lambda_{\textsc{b}}\int\dd s\dd s'\dd^{3}\bm{x}\dd^{3}\bm{x}'\,
\Theta(s-s') F_{\textsc{a}}(\bm{x})F_{\textsc{b}}(\bm{x}')\nonumber\\
&\quad\times\bigg\{
e^{\ii\Omega_{\textsc{a}}s+\ii\Omega_{\textsc{b}}s'} \chi_{\textsc{a}}(s)\chi_{\textsc{b}}(s')
\left<\vphantom{\phi}\right.\hspace{-0.6mm}\hat{\phi}(s,\bm{x})\hat{\phi}(s',\bm{x}')\left.\hspace{-0.6mm}\vphantom{\phi}\right>\nonumber\\
&\quad+e^{\ii\Omega_{\textsc{b}}s+\ii\Omega_{\textsc{a}}s'} \chi_{\textsc{b}}(s)\chi_{\textsc{a}}(s')
\left<\vphantom{\phi}\right.\hspace{-0.6mm}\hat{\phi}(s,\bm{x}')\hat{\phi}(s',\bm{x})\left.\hspace{-0.6mm}\vphantom{\phi}\right>
\bigg\}.\label{eqta18}
\end{align}
Intuitively, $\mathcal{L}_{ij}$ is said to quantify detector responses to local noise, while $\mathcal{M}$ quantifies detector responses to nonlocal correlations and entanglement~\cite{PhysRevD.95.105009}. Note that $\mathcal{M}$ consists of time-ordered two-point correlation functions. As mentioned above in \S\ref{sec2b} this propagator will be modified by the covariant UV cutoff, changing the dynamics of the entanglement harvesting protocol.

From the reduced density matrix, we can evaluate the entanglement negativity $\mathcal{N}=\operatorname{max}(0,\mathcal{N}^{(2)})$, where~\cite{PhysRevA.71.042104}
\begin{align}
\mathcal{N}^{(2)}&=-\frac{1}{2}\left(
\mathcal{L}_{\textsc{aa}}+\mathcal{L}_{\textsc{bb}}
-\sqrt{\left(\mathcal{L}_{\textsc{aa}}-\mathcal{L}_{\textsc{bb}}\right)^{2}+4\abs{\mathcal{M}}^{2}}\right).\label{eq16n}
\end{align}
The entanglement negativity serves as an indicator of distillable entanglement. Additionally, in the case of two qubits (as we have here), the negativity is non-zero if and only if the qubits are entangled~\cite{PhysRevLett.77.1413}, allowing us to use negativity as an entanglement measure. 

A quick study of \eqref{eq16n} illustrates the general behavior of entanglement harvesting. To maximize negativity the local noise term $\mathcal{L}_{ii}$ needs to be suppressed with respect to the nonlocal term $\mathcal{M}$. Increasing the detectors' energy gap $\Omega_{i}$ is one way of suppressing local noise~\cite{PhysRevD.95.105009}. Note that the purpose of \eqref{eq16n} is to quantify quantum entanglement in a field, not to reveal the causal structure of the field. Consequently, it is common to use detectors whose spatial ($F_{i}(\bm{x})$) and temporal ($\chi_{i}(t)$) profiles that are not necessarily compactly supported. The most common of these profiles is a Gaussian function, due to its exponential suppression of nonlocal tails, and useful analytic and integration properties.

\subsection{Harvesting with covariant UV cutoffs}

\begin{figure}[!t]
\includegraphics[width=0.9\columnwidth]{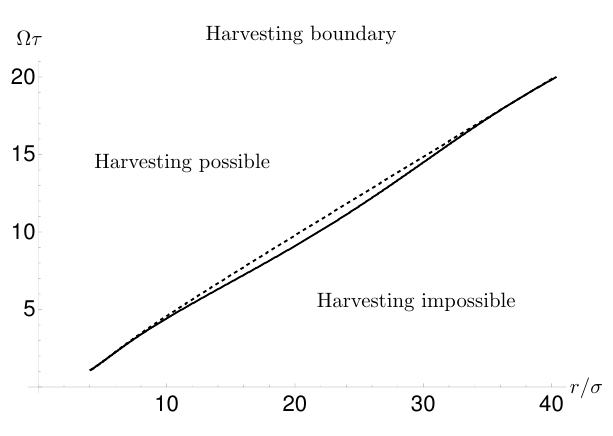}
\caption{\label{fig1} Harvesting boundary for identical detectors spaced a distance~$r$ apart. The solid line has a cutoff~$\Lambda=0.1/\tau$, and the dotted line has $\Lambda=10/\tau$. The boundary for $\Lambda>10/\tau$ is indistinguishable from the $\Lambda=10/\tau$ and is therefore omitted. The covariant UV cutoff has increased the harvesting range, although this increase in range oscillates with $\Omega$. The small distance $r/\sigma<4$ has been omitted as this involves significant detector smearing $F_{i}(\bm{x})$ overlap and corresponds to a physical scenario that is poorly described by the UDW detector model.}
\end{figure}

For computational simplicity, we use Gaussian switching and smearing functions and consider both detectors to have the same energy gap, switching function, and smearing function (up to a spatial translation): 
\begin{align}
\chi_{\textsc{d}}(t)&=\frac{1}{\sqrt{2\pi\tau^{2}}}e^{-\frac{t^{2}}{2\tau^{2}}},\\
F_{\textsc{d}}(\bm{x})&=\frac{1}{(2\pi\sigma^{2})^{3/2}}e^{-\frac{\abs{\bm{x}-\bm{\mu}_{\textsc{d}}}^{2}}{2\sigma^{2}}}.
\end{align}
In flat-spacetime scenarios, negativity tends to decrease with distance up to a well-defined boundary, after which harvesting is impossible, and increasing the detectors' energy gaps tends to increase this maximum range. Therefore, the binary and well-defined nature of the harvesting boundary is useful in illustrating the effects of the covariant UV cutoff. In figure~\ref{fig1}, we have plotted the boundary of possible entanglement harvesting as a function of the detector energy gap~$\Omega$. The harvesting boundary $(r,\Omega)$ is plotted for a small UV cutoff~$\Lambda=0.1/\tau$ and a large UV cutoff $\Lambda=2.5/\tau$. Plots with even larger $\Lambda$ were omitted as they overlap indiscriminately with $\Lambda=2.5/\tau$.

If we consider $\Omega$ as the independent variable in figure~\ref{fig1}, we can see a measurable effect of the covariant bandlimit via changes in the theoretical range of successful entanglement harvesting. The generally monotonic growth of both plots in figure~\ref{fig1} is unsurprising. The changes between the non-cutoff and cutoff plots show the expected increase in the harvesting range. The small $r/\sigma<4$ behavior is omitted as this describes a scenario with significant detector ($F_{i}(\bm{x})$) overlap. This represents a physical scenario that is poorly described by the UDW model and would require Pauli exclusion or molecular considerations.

From \S\ref{sec3a}, we know that introducing a covariant cutoff introduces acausal behavior from the modified virtual particle channels, and this effect is also seen here with an increase in the range of allowable harvesting. The similarities of \eqref{eq5} and \eqref{eqta18} demonstrate that the increase in harvesting range arises from an acausal communication channel between the two detectors, mediated by virtual particle excitations. This strengthens the nonlocal correlations of the two detectors (quantified by $\abs{\mathcal{M}}$) while leaving the local noise unchanged ($\mathcal{L}_{ij}$). This particular situation highlights a major difference between covariant and non-covariant bandlimitation, which is that covariant bandlimitation modifies two-detector interactions in a non-trivially different manner to single-detector interactions. This means we cannot na\"{i}vely interpret a covariant UV cutoff as equivalent to a modification of the detector's spatial profile as is the case in non-covariant UV cutoffs~\cite{PhysRevD.102.125026}.

\begin{figure}[!t]
\includegraphics[width=1.0\columnwidth]{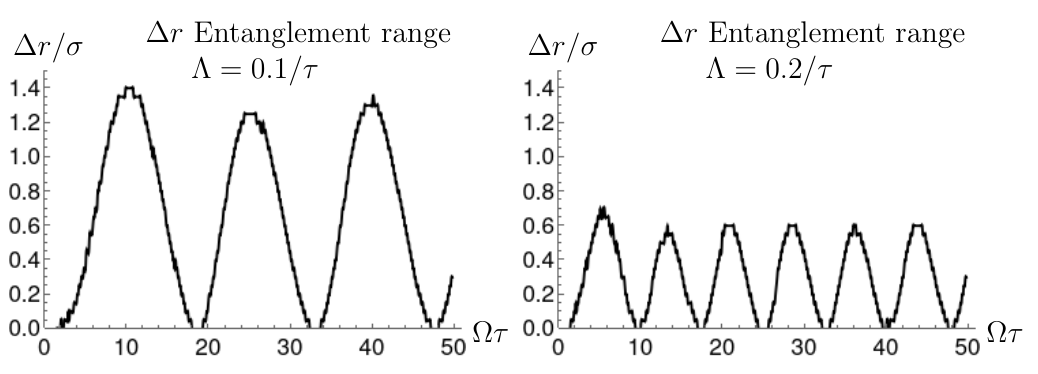}
\caption{\label{fig2} Differences in harvesting range for $\Lambda=0.1/\tau$ and $\Lambda=0.2/\tau$ vs $\Lambda\rightarrow \infty$. Despite the oscillatory behavior, the difference is always positive. Note how the frequency of oscillation increases and the amplitude decreases with $\Lambda$. The small~$\Omega$ behavior has been omitted as this corresponds to overlapping detector smearings where the UDW model is a poor description. Note the sharp edges of these plots are numerical artifacts due to low plot resolution. }
\end{figure}

An inspection of the plots in figure~\ref{fig1} shows that the increase in harvesting range quantitatively varies with $\Omega$. These variations are plotted in figure~\ref{fig2} and show the oscillatory nature of these variations. Curiously, while the entanglement range oscillates with changing $\Omega$ it does not become negative. Figure~\ref{fig2} can be interpreted as illustrating a resonance between the detectors and the virtual particles propagating between them. The $\Omega$ values where the entanglement range variations are zero seem to be those where the detectors are insensitive to the causality-violating virtual particles. However, it should be noted that this interpretation is a loose analogy and should not be abused.

\begin{figure}[!t]
\includegraphics[width=0.9\columnwidth]{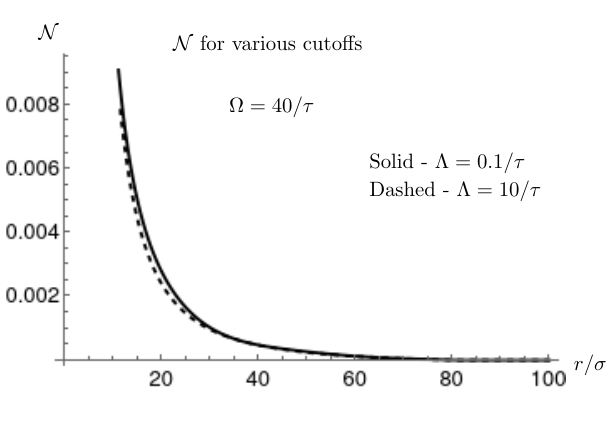}
\caption{\label{fig3} Negativity vs detector separation for $\Omega=40/\tau$. Despite the slight increase in entanglement range, the quantity of negativity is mostly unaffected by the cutoff. That is, the entangling power of the virtual particles is quite limited.}
\end{figure}

In figure~\ref{fig3}, we plot the entanglement negativity for an energy gap $\Omega=40/\tau$ for two different cutoffs, $\Lambda=0.1/\tau$ and $\Lambda=10/\tau$. We already know there will be a shift in the entanglement range, although it is not evident from figure~\ref{fig3}. In figure~\ref{fig3}, the relative difference between the two plots shows that the entanglement harvested by the detectors is relatively insensitive to the scale of the covariant cutoff. We can interpret this as telling us that the acausal communication introduced by the covariant UV cutoff only weakly contributes to the entanglement harvested. Specifically, considering figure~\ref{fig4}, smooth detectors (Gaussian profiles) are weakly impacted by the acausal field behavior and only detect minor increases in nonlocal correlations (encoded by $\mathcal{M}$). From the results of two-qubit communication and entanglement harvesting we should visualize the acausal behavior of the covariant UV cutoff model as blurring the lightcone on the scale of the minimum length and expect this behavior to be almost imperceptible with inertial detectors of atomic size.

\section{Discussion}\label{sec_z5}


\subsection{Nonlocality and signaling}
The explicit use of a UV cutoff without further modifications to the field theory introduces acausal behavior for both non-covariant~\cite{PhysRevD.92.104019} and covariant bandlimitation. The nature of the acausalities differ in the two cases, with non-covariantly bandlimited fields often interpreted as a loss in detector locality~\cite{PhysRevD.102.125026,PhysRevD.108.096024}. Conversely, the covariant bandlimited field we are studying here cannot be easily interpreted as a nonlocal detector theory but does lend itself to a natural interpretation as a field theory with acausal virtual particle excitations. 

As shown in \S\ref{sec4a}, the covariant cutoff acted trivially on the local terms $\mathcal{L}_{ij}$ and non-trivially on the nonlocal term $\mathcal{M}$. This uneven modification highlights a major difference with the non-covariant cutoff models, principally, that interpreting the covariant cutoff model as a nonlocal detector theory will not be as straightforward as in the non-covariant case. Considering the covariant cutoff model from a Green's function perspective, there may exist a detector nonlocality interpretation whose detector nonlocality will be entangled with the detector's internal degrees of freedom.

The complexity of the detector nonlocality raises issues when using detectors (e.g.,~UDW detectors~\cite{UDW_1,UDW_2}) as sources of field perturbations. Specifically, does a field excitation or pulse generated by some local detectors (e.g.,~a lasing crystal) causally propagate through a field with a covariant UV cutoff? Direct calculations of field expectation values (e.g.,~$\left<\vphantom{\phi}\right.\hspace{-0.6mm}\hat{\phi}^{2}(\bm{x},t)\left.\hspace{-0.6mm}\vphantom{\phi}\right>$ or $\left<\vphantom{\phi}\right.\hspace{-0.6mm}\hat{T}_{\mu\nu}(\bm{x},t)\left.\hspace{-0.6mm}\vphantom{\phi}\right>$) suggest the pulse propagates causally. However, these field expectation values are insensitive to virtual particle excitations, which we know are the generators of acausal behavior~\S\ref{sec3a}.

The apparent transparency of virtual particles in field expectation values suggests that the algebra of observables~\cite{fewster2019algebraicquantumfieldtheory,PhysRevD.106.045012} in a covariantly cutoff field theory needs to be larger than the set of field operators. A possible expansion involves introducing a time-ordering operator into field expectation values. This would be trivial in non-cutoff theories but would introduce sensitivity to virtual particles in the covariantly cutoff field. 

Despite the complexity of the detector nonlocality, we do know that the acausal behavior occurs mostly near the lightcone boundary, i.e. the covariant cutoff blurs the lightcone boundary. The acausal behavior at large space-like separations is suppressed by the detector's smooth profiles that are large w.r.t. the minimum length of the model. It is also worth remembering that these acausal violations about the lightcone boundary are Lorentz invariant, creating fuzziness in space and time.

\subsection{Real vs virtual particles in covariantly cutoff QFTs}
This paper has focused on two detector protocols as they are the simplest 2nd-order perturbative protocols that exhibit nontrivial changes under a covariant UV cutoff. This follows since the entanglement harvesting and communications protocols are sensitive to the spectrum of virtual particles and have sufficient detector degrees of freedom to detect the changes introduced by the covariant bandlimit. 

As mentioned in the previous section, virtual particles introduce acausal behavior that is physically measurable. This has interesting consequences when studying QFTs on curved spacetimes with backreaction. For example, the simplest QFT gravitational backreaction model modifies Einstein's equation by replacing the classical stress-energy tensor with the expectation value of the QFT stress-energy tensor~\cite{Birrell1984}. This means that classical spacetime only uses real particles to source the stress-energy tensor and therefore responds causally to perturbations on the QFT. However, a more general interaction is expected to couple the field degrees of freedom with the metric, i.e. a generalization of the UDW interaction above (\S\ref{seca3a}). This would allow the metric to interact with the acausal virtual particle excitations and introduce non-trivial behavior in regions devoid of real particles. Na\"{i}vely appears to be a breakdown of strong local passivity~\cite{PhysRevE.90.012127,PhysRevLett.123.190601} in regions that are devoid of real particles and are locally vacuum (according to the expectation value of local field operators). This contradiction could be used as a constraint for possible interactions between gravity and quantum fields and is an open problem emerging from this work.

The changes to real and virtual particle behavior introduced by a covariant UV cutoff on a QFT have many additional effects. In addition to the example described above, we also have the following considerations.
\begin{itemize}
    \item \textit{Gravitational wave propagation}: If the spacetime metric interacts with a covariantly cutoff QFT, will the acausal QFT affect gravitational wave propagation? Could the virtual particle field act as a $n<1$ refractive medium?
    \item \textit{Virtual particles as information carriers}: If the acausal virtual particles do indeed carry information, What are the entropic consequences of this, given they do not carry any apparent energy (in the form of $\left<\vphantom{\phi}\right.\hspace{-0.6mm}\hat{T}_{\mu\nu}(\bm{x},t)\left.\hspace{-0.6mm}\vphantom{\phi}\right>$)? Also, how do acausal virtual particles influence the black-hole-information problem?
    \item \textit{Strong local passivity}: The acausal communication via virtual particles would allow a space-like separated quantum energy teleportation (QET) \cite{PhysRevD.96.025014} protocol to extract energy from a local field vacuum. How should strong local passivity be defined in a covariantly bandlimited field?
    \item \textit{Energy conditions}: How are quantum energy conditions~\cite{Ford78,Pfenning98} affected by a covariant UV cutoff?
    \item \textit{Acausal communication}: Can the acausal communication be exploited into new RQI protocols? For example, could acausal communication in QET be used to create the stress-energy tensor for a stable warp bubble metric~\cite{Alcubierre_1994,Barcelo2022}?
    \item \textit{Harmonic-oscillator detectors}: Harmonic oscillator UDW detectors~\cite{Brown_2014} can be modeled non-perturbatively, thereby obscuring the Feynman propagator and the effects of the covariant UV cutoff. Can we introduce an effective action term to account for the cutoff effects in non-perturbative models?
    \item \textit{Noether's theorem}: Do Noether's conserved quantities~\cite{Noether_thm} remain unchanged in a covariantly UV-cutoff QFT? How will the effective action mentioned above affect the Noether conserved quantities? Would these conserved quantities include contributions from virtual particles?
\end{itemize}

\subsection{Analytic observations}
The work presented above admits several interpretations for the changes introduced by a covariant UV cutoff. The previous discussion section focused on the effects of the covariant UV cutoff on virtual particle excitations and the causality violations it introduced. A study of \eqref{eqa5} suggests another interpretation of the covariant UV cutoff's effects, independent of propagators and Feynman diagram nomenclature. 

The unitary time evolution operator arises from the time-evolution axiom of quantum theory~\cite{liboff2003introductory} and is formally written as a time-ordered exponential of the interaction Hamiltonian, which is then Taylor expanded into a Dyson series~\cite{liboff2003introductory}. The covariant UV cutoff introduces modifications to the Heaviside step functions in the Dyson series and effectively modifies the action of the time-ordering operator~\cite{Pye_thesis}. This change is unlike any non-covariant UV cutoff or approximation applied to QFT, whose changes usually involve a truncation in the working Hilbert space or a modification to the Hamiltonian's form. Instead, as a result of the covariant UV cutoff, the Hamiltonian is no longer the exact generator of time translations, i.e.,~the Hermitian operator that generates the correct unitary operation, e.g.,~\eqref{eqa12}, with the cutoff in place. This appears to be a manifestation of the detector nonlocality, which importantly results in nonlocalities in time and space.

When considering the mathematical changes to the Dyson series, the covariant UV cutoff can be interpreted as introducing uncertainty in time---and thus a softening of the usually strict notion of time ordering. This means either the Dyson series is inappropriate for describing the time evolution operator (violation of the quantum time-evolution axiom) or time cannot be treated as a classical independent variable, and the fluctuations in time result in a local breakdown of causal order (c.f. indefinite-causal-order models~\cite{hardy2005probability, Oreshkov_2016}).

\section{Conclusion}\label{sec_z6}
In this manuscript, we have studied a quantum field subject to a covariant UV cutoff and its consequences in RQI protocols, namely communication and entanglement harvesting. As with the non-covariantly UV-cutoff theories, we found that introducing a covariant UV cutoff introduces acausal and nonlocal behavior (which is nontrivial only near the minimum length scale). Our study of the simple two-qubit communication protocol results in qualitative similarities between the two types of UV cutoffs, namely an oscillatory and polynomially decaying acausality. In both cases, the oscillation's frequency is of the same order as the UV cutoff and therefore tends to suppress extreme space-like communication when using physical detectors (i.e.,~smooth and large relative to the UV cutoff), resulting in an enlargement and blurring of the detector’s smearing on the order of the minimum length scale. Unlike the non-covariant case~\cite{PhysRevD.102.125026}, wherein the blurring is only in space, in the covariant case, this blurring also exists in time.

Our results from entanglement harvesting under a covariant UV cutoff quantitatively illustrate the effects of such a cutoff on physical observables. From a coarse qualitative perspective, covariant and non-covariant UV cutoff models have similar characteristics, namely an increase in the range of entanglement harvesting. Quantitatively, the covariant UV cutoff introduces minor changes to the entanglement negativity for intermediate detector separations (smaller than the harvesting boundary). This suggests that the acausal communication introduced by the cutoff contributes a relatively small amount (but non-zero) to the nonlocal correlation ($\mathcal{M}$) relative to the vacuum fluctuations.  

Our analytical study of the RQI protocols above demonstrates the incompleteness of our current picture of a covariantly cutoff (interacting) QFT with two major observations. Firstly, the covariant cutoff's influence on virtual particles means na\"{i}ve micro-causality no longer implies causal independence. Secondly, expectation values of field observables seem insensitive to the acausal virtual particle excitations. This means that full knowledge of all field expectation values is now informationally incomplete---i.e., insufficient to predict physical detector responses. Consequently, a covariantly cutoff QFT would require measurements capable of describing virtual particle distributions, and assuming we can find a correct effective action or temporally nonlocal Hamiltonian for this theory, it would also require a modification to field projective measurements and any Noether conserved quantities (e.g.,~virtual particle contributions to energy density).

\begin{acknowledgments}
We thank Achim Kempf, Jason Pye, Dominic Lewis, Gavin Brennen and Danny Terno, for useful discussions. This work was supported by the Australian Research Council Centre of Excellence for Quantum Computation and Communication Technology (Project No.~CE170100012) and the Australian Research Council Discovery Program (Project No.~DP200102152). 
\end{acknowledgments}

\onecolumngrid

\appendix

\section{Covariant UV cutoffs}\label{seca_appendix1}
\subsection{Mathematics of covariant UV cutoffs}\label{sec_a1}
Covariant bandlimitation involves truncating the spectrum of the d'Alembertian operator and ensuring that all field-related quantities and attributes are confined to the eigenspace defined by this restricted spectrum~\cite{Pye_thesis}. Consequences of this cutoff in cosmological settings are explored in Refs.~\cite{PhysRevLett.92.221301, 10.1063/1.4790482, PhysRevLett.119.031301}. A clear and useful review of covariant bandlimitation can be found in Jason Pye's PhD thesis~\cite{Pye_thesis}, including a demonstration of the equivalence of path-integral quantization and the canonical quantization in the context of covariant bandlimitation. Here we review some specific details from Pye's thesis.

In the main text, we discuss detector-field interactions using 2nd-order perturbation theory. The main tool used in this circumstance is the bandlimited projector
\begin{align}
    \Pi_{\Lambda} : L_2(\mathbb R^{3+1}) &\to L_2(\mathbb R^{3+1})
\nonumber \\
    f &\mapsto \Pi_{\Lambda}[f],
\intertext{where}
    \Pi_{\Lambda}[x \mapsto e^{\ii k_{\mu}x^{\mu}}]
&\coloneqq
    [
    x \mapsto 
    e^{\ii k_{\mu}x^{\mu}}
    \Theta(\Lambda^{2}-\abs{k_{\nu}k^{\nu}})
    ],
\intertext{and thus, for any function $f$,}
    \Pi_{\Lambda}[f](x)
&\coloneqq
   \int\limits_{\mathcal{B}}\frac{\dd^{4}k}{(2\pi)^{4}}e^{\ii k_{\mu}x^{\mu}}
   \tilde f(k)
,
\intertext{where $\tilde f$ is the Fourier transform of $f$, namely}
    \tilde f(k)
&\coloneqq
    \int\limits_{\mathbb{R}^{3+1}} \dd^4 y\, e^{-\ii k_{\mu}y^{\mu}} f(y),
\end{align}
and $\mathcal{B}=\{k_{\mu}\in\mathbb{R}^{3+1}:\abs{k_{\mu}k^{\mu}}<\Lambda^{2}\}$. 
Notice that $\Pi_\Lambda$ leaves plane waves in~$x$ unchanged provided their 4-momentum is within the bandlimit (and maps to the zero function otherwise). It is a projector because its eigenvalues are either 0 or 1 (the outputs of the Heaviside theta function), but because the projection takes place in terms of plane waves, it acts as a convolution in real space:
\begin{align}
    \Pi_{\Lambda}[f](x)
&=
    \int\limits_{\mathbb{R}^{3+1}}
    \frac{\dd^{4}k}{(2\pi)^{4}}
    e^{\ii k_{\mu}x^{\mu}}
    \Theta(\Lambda^{2}-\abs{k_{\nu}k^{\nu}})
    \int\limits_{\mathbb{R}^{3+1}} \dd^4 y\,
    e^{-\ii k_{\mu}y^{\mu}} f(y)
\nonumber \\
&=
    \int\limits_{\mathbb{R}^{3+1}} \dd^4 y\,
     f(y)
    \int\limits_{\mathbb{R}^{3+1}}
    \frac{\dd^{4}k}{(2\pi)^{4}}
    e^{\ii k_{\mu}(x-y)^{\mu}}
    \Theta(\Lambda^{2}-\abs{k_{\nu}k^{\nu}})
\nonumber \\
&=
    \int\limits_{\mathbb{R}^{3+1}} \dd^4 y\,
     f(y)
     \delta_\Lambda(x-y)
\nonumber \\
&=
    [f * \delta_\Lambda](x)
    ,
\intertext{where}
    \delta_\Lambda(x)
&\coloneqq
    \int\limits_{\mathcal{B}}
    \frac{\dd^{4}k}{(2\pi)^{4}}
    e^{\ii k_{\mu}x^{\mu}},\label{eqaa6}
\end{align}
and $*$ is 3+1~D convolution. Similarly to the sinc function in non-covariant UV cutoff~\cite{PhysRevD.102.125026}, $\delta_{\Lambda}$ is an indicator of the acausal nonlocality introduced by the covariant UV cutoff. This allows us to define
\begin{align}
    \Pi_{\Lambda} : L_2(\mathbb R^{3+1}) &\to L_2(\mathbb R^{3+1})
\nonumber \\
    f &\mapsto f * \delta_\Lambda.\label{eqaa7}
\end{align}
Also notice that, from its definition, $\delta_\Lambda(x) = \delta_\Lambda(x)^*$, which means that if $f$ is a real-valued function, then so is $\Pi_\Lambda[f]$.

In a perturbative expansion, these projectors act on the spacetime variables associated with a Wick contraction. In 2nd-order perturbation theory, this is usually implicitly performed when tracing out the field degrees of freedom. For higher-order perturbative expansions, it is easier to explicitly use Wick's theorem and apply the $\Pi_{\Lambda}$ projectors on the Feynman propagators resulting from the Wick contractions. 

The mathematics of the covariant bandlimit explored in this appendix is subject to ambiguity concerning its application to quantum field theory. It is important to note that the covariant bandlimitation method used in this paper and Pye's thesis~\cite{Pye_thesis} has been specifically chosen to ensure the equivalence of canonical quantization and path-integral quantization. The action of the $\Pi_{\Lambda}$ projectors on the field propagators is equivalent to the unambiguously defined covariant UV cutoff in path-integral quantization of truncating the functional integral's integration domain.

\subsection{Dyson expansion under covariant cutoffs}\label{sec_a2}

In the linear UDW detector models we employ, the interaction Hamiltonian (in the interaction picture) takes the form
\begin{align}
\hat{H}_{\textsc{i}}&=\sum_{i\in\{\textsc{a,b}\}}\lambda_{i}\chi_{i}(t)\hat{\sigma}^{i}_{x}(t)\int\dd^{3}\bm{x}\,F_{i}(\bm{x})\hat{\phi}(t,\bm{x}),\label{eqaa8}
\end{align}
where $\lambda_{i}$ is the interaction strength, $\chi_{i}$ is the switching function (when the detector is on) and $F_{i}$ is the smearing (effective size of the detector). 

As with most RQI protocols, we use a 2nd-order perturbative expansion to model the system (taking initial and final times as $-\infty$ and $+\infty$ respectively):
\begin{align}
\hat{U}&=\mathcal{T}\text{exp}\left[-\ii\int\limits_{-\infty}^{\infty}\dd t\,\hat{H}(t)\right],\\
&=\mathbb{I}-\ii\int\limits_{-\infty}^{\infty}\dd s\,\hat{H}_{\textsc{i}}(s)-\int\hspace{-2mm}\int\limits_{\mathbb{R}^{2}}\dd s\,\dd s'\,\Theta(s-s')\hat{H}_{\textsc{i}}(s)\hat{H}_{\textsc{i}}(s')+\mathcal{O}(\lambda^{3}).
\end{align}
For further notational simplicity let
\begin{align}
\hat{\Phi}_{i}(t)&=\lambda_{i}\chi_{i}(t)\int\dd^{3}\bm{x}\,F_{i}(\bm{x})\hat{\phi}(t,\bm{x}),
\end{align}
such that
\begin{align}
\hat{H}_{\textsc{i}}&=\hat{\sigma}_{x}^{\textsc{a}}(t)\hat{\Phi}_{\textsc{a}}(t)+\hat{\sigma}_{x}^{\textsc{b}}(t)\hat{\Phi}_{\textsc{b}}(t).
\end{align}
We consider the commonly used initial conditions $\hat{\rho}_{0}=\hat{\rho}_{d,0}\otimes\hat{\rho}_{\phi}$, i.e. the detectors and field are initially separable. Then the post-interaction density matrix becomes

\begin{align}
\hat{\rho}&=\hat{U}\hat{\rho}_{0}\hat{U}^{\dagger}.
\end{align}

Most RQI protocols focus on the detector responses so we trace out the field degree of freedom:

\begin{align}
\hat{\rho}_{d,f}&=\Tr_{\phi}\hat{\rho},\\
\begin{split}
&=\Tr_{\phi}\left[
\left(\mathbb{I}-\ii\int\limits_{-\infty}^{\infty}\dd s\,\hat{H}_{\textsc{i}}(s)-\int\hspace{-2mm}\int\limits_{\mathbb{R}^{2}}\dd s\,\dd s'\,\Theta(s-s')\hat{H}_{\textsc{i}}(s)\hat{H}_{\textsc{i}}(s')\right)\right.\hat{\rho}_{0}\\
&\quad\times\left.\left(\mathbb{I}+\ii\int\limits_{-\infty}^{\infty}\dd s\,\hat{H}_{\textsc{i}}(s)-\int\hspace{-2mm}\int\limits_{\mathbb{R}^{2}}\dd s\,\dd s'\,\Theta(s-s')\hat{H}_{\textsc{i}}(s')\hat{H}_{\textsc{i}}(s)\right)\right],
\end{split}\\
\begin{split}
&=\Tr_{\phi}\left[
\vphantom{\left(\mathbb{I}-\ii\int\limits_{-\infty}^{\infty}\dd s\,\hat{H}_{\textsc{i}}(s)-\int\hspace{-2mm}\int\limits_{\mathbb{R}^{2}}\dd s\,\dd s'\,\Theta(s-s')\hat{H}_{\textsc{i}}(s)\hat{H}_{\textsc{i}}(s')\right)}\right.
\hat{\rho}_{0}-\ii\int\limits_{-\infty}^{\infty}\dd s\,\left(
\left[
\hat{\sigma}_{x}^{\textsc{a}}(s)\hat{\Phi}_{\textsc{a}}(s)+\hat{\sigma}_{x}^{\textsc{b}}(s)\hat{\Phi}_{\textsc{b}}(s)
\right]\hat{\rho}_{0}
-\hat{\rho}_{0}\left[
\hat{\sigma}_{x}^{\textsc{a}}(s)\hat{\Phi}_{\textsc{a}}(s)+\hat{\sigma}_{x}^{\textsc{b}}(s)\hat{\Phi}_{\textsc{b}}(s)
\right]\right)\\
&\quad + \int\hspace{-2mm}\int\limits_{\mathbb{R}^{2}} \dd s\dd s'
\left[
\hat{\sigma}_{x}^{\textsc{a}}(s)\hat{\Phi}_{\textsc{a}}(s)+\hat{\sigma}_{x}^{\textsc{b}}(s)\hat{\Phi}_{\textsc{b}}(s)
\right]\hat{\rho}_{0}
\left[
\hat{\sigma}_{x}^{\textsc{a}}(s')\hat{\Phi}_{\textsc{a}}(s')+\hat{\sigma}_{x}^{\textsc{b}}(s')\hat{\Phi}_{\textsc{b}}(s')
\right]\\
&\quad -\int\hspace{-2mm}\int\limits_{\mathbb{R}^{2}} \dd s\dd s'\,\Theta(s-s')\left(
\left[
\hat{\sigma}_{x}^{\textsc{a}}(s)\hat{\Phi}_{\textsc{a}}(s)+\hat{\sigma}_{x}^{\textsc{b}}(s)\hat{\Phi}_{\textsc{b}}(s)
\right]
\left[
\hat{\sigma}_{x}^{\textsc{a}}(s')\hat{\Phi}_{\textsc{a}}(s')+\hat{\sigma}_{x}^{\textsc{b}}(s')\hat{\Phi}_{\textsc{b}}(s')
\right]\hat{\rho}_{0}\right)\\
&\quad-\int\hspace{-2mm}\int\limits_{\mathbb{R}^{2}} \dd s\dd s'\,\Theta(s-s')\left(\hat{\rho}_{0}
\left[
\hat{\sigma}_{x}^{\textsc{a}}(s')\hat{\Phi}_{\textsc{a}}(s')+\hat{\sigma}_{x}^{\textsc{b}}(s')\hat{\Phi}_{\textsc{b}}(s')
\right]\left[
\hat{\sigma}_{x}^{\textsc{a}}(s)\hat{\Phi}_{\textsc{a}}(s)+\hat{\sigma}_{x}^{\textsc{b}}(s)\hat{\Phi}_{\textsc{b}}(s)
\right]\right)+\mathcal{O}(\lambda^{3})
\left.
\vphantom{\left(\mathbb{I}-\ii\int\limits_{-\infty}^{\infty}\dd s\,\hat{H}_{\textsc{i}}(s)-\int\hspace{-2mm}\int\limits_{\mathbb{R}^{2}}\dd s\,\dd s'\,\Theta(s-s')\hat{H}_{\textsc{i}}(s)\hat{H}_{\textsc{i}}(s')\right)}\right],
\end{split}\\
\begin{split}
&=
\hat{\rho}_{d,0}-\ii\int\limits_{-\infty}^{\infty}\dd s\,\left(
\hat{\sigma}_{x}^{\textsc{a}}(s)\hat{\rho}_{d,0} \left<\vphantom{\phi}\right.\hspace{-0.6mm}\hat{\Phi}_{\textsc{a}}(s)\left.\hspace{-0.6mm}\vphantom{\phi}\right>+
\hat{\sigma}_{x}^{\textsc{b}}(s)\hat{\rho}_{d,0} \left<\vphantom{\phi}\right.\hspace{-0.6mm}\hat{\Phi}_{\textsc{b}}(s)\left.\hspace{-0.6mm}\vphantom{\phi}\right>
-\hat{\rho}_{d,0}\hat{\sigma}_{x}^{\textsc{a}}(s) \left<\vphantom{\phi}\right.\hspace{-0.6mm}\hat{\Phi}_{\textsc{a}}(s)\left.\hspace{-0.6mm}\vphantom{\phi}\right>-
\hat{\rho}_{d,0}\hat{\sigma}_{x}^{\textsc{b}}(s) \left<\vphantom{\phi}\right.\hspace{-0.6mm}\hat{\Phi}_{\textsc{b}}(s)\left.\hspace{-0.6mm}\vphantom{\phi}\right>\right)\\
&\quad+\int\hspace{-2mm}\int\limits_{\mathbb{R}^{2}} \dd s\dd s' \left(
 \hat{\sigma}_{x}^{\textsc{a}}(s)\hat{\rho}_{d,0}\hat{\sigma}_{x}^{\textsc{a}}(s') \left<\vphantom{\phi}\right.\hspace{-0.6mm}\hat{\Phi}_{\textsc{a}}(s')\hat{\Phi}_{\textsc{a}}(s)\left.\hspace{-0.6mm}\vphantom{\phi}\right>
+\hat{\sigma}_{x}^{\textsc{a}}(s)\hat{\rho}_{d,0}\hat{\sigma}_{x}^{\textsc{b}}(s') \left<\vphantom{\phi}\right.\hspace{-0.6mm}\hat{\Phi}_{\textsc{a}}(s')\hat{\Phi}_{\textsc{b}}(s)\left.\hspace{-0.6mm}\vphantom{\phi}\right>\right.\\
&\quad\left.+\hat{\sigma}_{x}^{\textsc{b}}(s)\hat{\rho}_{d,0}\hat{\sigma}_{x}^{\textsc{a}}(s') \left<\vphantom{\phi}\right.\hspace{-0.6mm}\hat{\Phi}_{\textsc{b}}(s')\hat{\Phi}_{\textsc{a}}(s)\left.\hspace{-0.6mm}\vphantom{\phi}\right>
+\hat{\sigma}_{x}^{\textsc{b}}(s)\hat{\rho}_{d,0}\hat{\sigma}_{x}^{\textsc{b}}(s') \left<\vphantom{\phi}\right.\hspace{-0.6mm}\hat{\Phi}_{\textsc{b}}(s')\hat{\Phi}_{\textsc{b}}(s)\left.\hspace{-0.6mm}\vphantom{\phi}\right>
\right)\\
&\quad -\int\hspace{-2mm}\int\limits_{\mathbb{R}^{2}} \dd s\dd s'\,\Theta(s-s')\left(
\left[
 \hat{\sigma}_{x}^{\textsc{a}}(s)\hat{\sigma}_{x}^{\textsc{a}}(s')
\left<\vphantom{\phi}\right.\hspace{-0.6mm}\hat{\Phi}_{\textsc{a}}(s)\hat{\Phi}_{\textsc{a}}(s')\left.\hspace{-0.6mm}\vphantom{\phi}\right>
+\hat{\sigma}_{x}^{\textsc{a}}(s)\hat{\sigma}_{x}^{\textsc{b}}(s')
\left<\vphantom{\phi}\right.\hspace{-0.6mm}\hat{\Phi}_{\textsc{a}}(s)\hat{\Phi}_{\textsc{b}}(s')\left.\hspace{-0.6mm}\vphantom{\phi}\right>
+\hat{\sigma}_{x}^{\textsc{b}}(s)\hat{\sigma}_{x}^{\textsc{a}}(s')
\left<\vphantom{\phi}\right.\hspace{-0.6mm}\hat{\Phi}_{\textsc{b}}(s)\hat{\Phi}_{\textsc{a}}(s')\left.\hspace{-0.6mm}\vphantom{\phi}\right>\right.\right.\\
&\quad\left.+\hat{\sigma}_{x}^{\textsc{b}}(s)\hat{\sigma}_{x}^{\textsc{b}}(s')
\left<\vphantom{\phi}\right.\hspace{-0.6mm}\hat{\Phi}_{\textsc{b}}(s)\hat{\Phi}_{\textsc{b}}(s')\left.\hspace{-0.6mm}\vphantom{\phi}\right>\right]\hat{\rho}_{d,0}
+\hat{\rho}_{d,0}\left[
 \hat{\sigma}_{x}^{\textsc{a}}(s')\hat{\sigma}_{x}^{\textsc{a}}(s)
\left<\vphantom{\phi}\right.\hspace{-0.6mm}\hat{\Phi}_{\textsc{a}}(s')\hat{\Phi}_{\textsc{a}}(s)\left.\hspace{-0.6mm}\vphantom{\phi}\right>
+\hat{\sigma}_{x}^{\textsc{a}}(s')\hat{\sigma}_{x}^{\textsc{b}}(s)
\left<\vphantom{\phi}\right.\hspace{-0.6mm}\hat{\Phi}_{\textsc{a}}(s')\hat{\Phi}_{\textsc{b}}(s)\left.\hspace{-0.6mm}\vphantom{\phi}\right>\right.\\
&\quad\left.\left.+\hat{\sigma}_{x}^{\textsc{b}}(s')\hat{\sigma}_{x}^{\textsc{a}}(s)
\left<\vphantom{\phi}\right.\hspace{-0.6mm}\hat{\Phi}_{\textsc{b}}(s')\hat{\Phi}_{\textsc{a}}(s)\left.\hspace{-0.6mm}\vphantom{\phi}\right>
+\hat{\sigma}_{x}^{\textsc{b}}(s')\hat{\sigma}_{x}^{\textsc{b}}(s)
\left<\vphantom{\phi}\right.\hspace{-0.6mm}\hat{\Phi}_{\textsc{b}}(s')\hat{\Phi}_{\textsc{b}}(s)\left.\hspace{-0.6mm}\vphantom{\phi}\right>\right]\right).
\end{split}\label{eqa12}
\end{align}
where the expectation values are evaluated over the field's initial state (the vacuum, for our purposes). 

If we inspect the expression above, we have linear field expectation values, ordinary quadratic field expectation values, and time-ordered quadratic field expectation values. The first two are unaffected by the covariant UV cutoff (on-shell), while the time-ordered quadratic expectation values will experience a non-trivial modification. 

For simplicity, we focus on one of the terms above and define
\begin{align}
\mathcal{F}&\coloneqq\int\dd s\dd s'\,\Theta(s-s')\hat{\sigma}_{x}^{\textsc{a}}(s)\hat{\sigma}_{x}^{\textsc{b}}(s')\hat{\rho}_{d,0}\left<\vphantom{\phi}\right.\hspace{-0.6mm}\hat{\Phi}_{\textsc{a}}(s)\hat{\Phi}_{\textsc{b}}(s')\left.\hspace{-0.6mm}\vphantom{\phi}\right>,\label{eqa18}\\
&=\lambda_{\textsc{a}}\lambda_{\textsc{b}}\int\dd s\dd s'\dd^{3}\bm{x}\dd^{3}\bm{x}'\,
\hat{\sigma}_{x}^{\textsc{a}}(s)\hat{\sigma}_{x}^{\textsc{b}}(s')\hat{\rho}_{d,0}
\chi_{\textsc{a}}(s)\chi_{\textsc{b}}(s')F_{\textsc{a}}(\bm{x})F_{\textsc{b}}(\bm{x}')
\Theta(s-s')\left<\vphantom{\phi}\right.\hspace{-0.6mm}\hat{\phi}(s,\bm{x})\hat{\phi}(s',\bm{x}')\left.\hspace{-0.6mm}\vphantom{\phi}\right>.\label{eqa19}
\end{align}
Note that $\mathcal{F}$ is a density matrix element with no spacetime dependence. To implement the covariant UV cutoff we need to consider the physical process that $\mathcal{F}$ describes, i.e. what is the Feynman diagram that corresponds to the amplitude $\mathcal{F}$. The RHS of equation \eqref{eqa18} contains the relevant information regarding the spacetime locations for the creation and annihilation of virtual particles and is explicit when expressed as the RHS of \eqref{eqa19}. 

Considering $\mathcal{F}$ alone provides no intuition on how to implement a covariant UV cutoff. It is therefore necessary to remember details such as the RHS of \eqref{eqa19} until the covariant UV cutoff has been implemented. This is the price to pay when not using path-integral quantization or explicit Green's function methods such as Wick's theorem.


Using the RHS of \eqref{eqa19}, the introduction of the covariant UV cutoff can be written as:
\begin{align}
\Pi_{\Lambda}\mathcal{F}&=\lambda_{\textsc{a}}\lambda_{\textsc{b}}\int\dd s\dd s'\dd^{3}\bm{x}\dd^{3}\bm{x}'\,
\hat{\sigma}_{x}^{\textsc{a}}(s)\hat{\sigma}_{x}^{\textsc{b}}(s')
\chi_{\textsc{a}}(s)\chi_{\textsc{b}}(s')F_{\textsc{a}}(\bm{x})F_{\textsc{b}}(\bm{x}')
\Pi_{\Lambda}\left[\Theta(s-s')\left<\vphantom{\phi}\right.\hspace{-0.6mm}\hat{\phi}(s,\bm{x})\hat{\phi}(s',\bm{x}')\left.\hspace{-0.6mm}\vphantom{\phi}\right>\right],\label{eqa9}
\end{align}
where we have implicitly modified the Green's function in the RHS of \eqref{eqa19}.

Using the Fourier representation of the Heaviside step function
\begin{align}
\Theta(t)&=\lim_{\epsilon\rightarrow 0^{+}}\frac{1}{2\pi\ii}\int\limits_{-\infty}^{\infty}\frac{\dd\nu}{\nu-\ii\epsilon}e^{\ii\nu t},
\end{align}
we can write out the Green's function in a Fourier basis. Recall that the initial field state is in the vacuum, then:
\begin{align}
\Theta(s-s')\left<\vphantom{\phi}\right.\hspace{-0.6mm}\hat{\phi}(s,\bm{x})\hat{\phi}(s',\bm{x}')\left.\hspace{-0.6mm}\vphantom{\phi}\right>&=\lim_{\epsilon\rightarrow 0^{+}}\frac{1}{2\pi\ii}\int\limits_{-\infty}^{\infty}\frac{\dd\nu}{\nu-\ii\epsilon}e^{\ii\nu(s-s')}\int\dd^{3}\bm{k}\dd^{3}\bm{k}'\,\frac{1}{\sqrt{4\omega\omega'}}\bra{0}
\left(\hat{a}_{\bm{k}}^{\vphantom{\dagger}}e^{-\ii\omega s+\ii\bm{k}\cdot\bm{x}}
+\hat{a}_{\bm{k}}^{\dagger}e^{\ii\omega s-\ii\bm{k}\cdot\bm{x}}\right)\nonumber\\
&\times
\left(\hat{a}_{\bm{k}'}^{\vphantom{\dagger}}e^{-\ii\omega' s'+\ii\bm{k}'\cdot\bm{x}'}
+\hat{a}_{\bm{k}'}^{\dagger}e^{\ii\omega' s'-\ii\bm{k}'\cdot\bm{x}'}\right)\ket{0},\\
&=\lim_{\epsilon\rightarrow 0^{+}}\frac{1}{2\pi\ii}\int\limits_{-\infty}^{\infty}\frac{\dd\nu}{\nu-\ii\epsilon}e^{\ii\nu(s-s')}\int\frac{\dd^{3}\bm{k}}{2\omega}e^{-\ii\omega s+\ii\bm{k}\cdot\bm{x}}
e^{\ii\omega s'-\ii\bm{k}\cdot\bm{x}'},\\
&=\lim_{\epsilon\rightarrow 0^{+}}\frac{1}{2\pi\ii}\int\limits_{-\infty}^{\infty}\frac{\dd\nu}{\nu-\ii\epsilon}\int\frac{\dd^{3}\bm{k}}{2\omega}e^{\ii\nu(s-s')} e^{-\ii\omega(s-s')+\ii\bm{k}\cdot(\bm{x}-\bm{x}')},\\
&=\lim_{\epsilon\rightarrow 0^{+}}\frac{1}{2\pi\ii}\int\limits_{-\infty}^{\infty}\frac{\dd\nu}{\nu-\ii\epsilon}\int\frac{\dd^{3}\bm{k}}{2\omega}e^{\ii(\nu-\omega)(s-s')+\ii\bm{k}\cdot(\bm{x}-\bm{x}')}.
\end{align}
If we now consider the action of the bandlimited projector on the expression above one can see that acting the projector with respect to  $(s,\bm{x})$, $(s',\bm{x}')$ or both leads to the same result. Therefore the covariant UV cutoff has the effect:
\begin{align}
\Pi_{\Lambda}\Theta(s-s')\left<\vphantom{\phi}\right.\hspace{-0.6mm}\hat{\phi}(s,\bm{x})\hat{\phi}(s',\bm{x}')\left.\hspace{-0.6mm}\vphantom{\phi}\right>&=\lim_{\epsilon\rightarrow 0^{+}}\frac{1}{2\pi\ii}\int\limits_{-\infty}^{\infty}\frac{\dd\nu}{\nu-\ii\epsilon}\int\frac{\dd^{3}\bm{k}}{2\omega}e^{\ii(\nu-\omega)t+\ii\bm{k}\cdot\bm{z}}\Theta\left(\Lambda^{2}-\abs{(\nu-\omega)^{2}-\abs{\bm{k}}^{2}}\right),
\end{align}
where we have defined $(t,\bm{z})=(s-s',\bm{x}-\bm{x}')$ for simplicity. Note that this can also be written as a convolution \eqref{eqaa7}
\begin{align}
\Theta(s-s')\left<\vphantom{\phi}\right.\hspace{-0.6mm}\hat{\phi}(s,\bm{x})\hat{\phi}(s',\bm{x}')\left.\hspace{-0.6mm}\vphantom{\phi}\right>&=\Theta(t)D^{+}(\Delta x),\\
\Pi_{\Lambda}[\Theta D^{+}](\Delta x)&=[(\Theta D^{+})*\delta_{\Lambda}](\Delta x),
\end{align}
where $\Delta x=(s-s',\bm{x}-\bm{x}')$ is a 4-vector, $t=s-s'$ and $D^{+}$ is the Wightman function (two-point field correlator). $\delta_{\Lambda}$ is the bandlimited Dirac delta function as defined in the previous section \eqref{eqaa6}.

Putting this together then means:
\begin{align}
\Pi_{\Lambda}\mathcal{F}&=\lambda_{a}\lambda_{b}\int\dd s\dd s'\dd^{3}\bm{x}\dd^{3}\bm{x}'\,
\hat{\sigma}_{x}^{a}(s)\hat{\sigma}_{x}^{b}(s')
\chi_{a}(s)\chi_{b}(s')F_{a}(\bm{x})F_{b}(\bm{x}')\nonumber\\
&\times\lim_{\epsilon\rightarrow 0^{+}}\frac{1}{2\pi\ii}\int\limits_{-\infty}^{\infty}\frac{\dd\nu}{\nu-\ii\epsilon}\int\frac{\dd^{3}\bm{k}}{2\omega}e^{\ii(\nu-\omega)t+\ii\bm{k}\cdot\bm{z}}\Theta\left(\Lambda^{2}-\abs{(\nu-\omega)^{2}-\abs{\bm{k}}^{2}}\right).\label{eqb18}
\end{align}
Note: the $\epsilon$ regularization describes the path of the integration contour about the $\nu=0$ singularity.

When considering \eqref{eqa12}, the covariant UV cutoff selectively changes some of the sharp, discontinuous Heaviside step functions into smoother, gradual steps. This can be interpreted as a modification to the time-ordering operator $\mathcal{T}$ so that it now allows time-disordered terms to contribute to unitary evolution. This is a major change to quantum mechanics and field theory that has always maintained the Dyson time-ordered form irrespective of the model. In some sense this can be interpreted as a blurring in time, i.e.,~time is no longer an absolute parameter (up to relativity) but now has some uncertainty that even Hamiltonian interactions cannot discern. 

\subsection{Commutation relations and micro-causality}\label{sec_a3}
Consider the commutatator
\begin{align}
\left[\hat{\phi}(t,\bm{x}),\hat{\phi}(t',\bm{x}')\right]&=\left[\int\frac{\dd^{3}\bm{k}}{(2\pi)^{3/2}}\frac{1}{\sqrt{2\omega}}\left(\hat{a}_{\bm{k}}^{\vphantom{\dagger}}e^{-\ii\omega t+\ii\bm{k}\cdot\bm{x}}+\hat{a}_{\bm{k}}^{\dagger}e^{\ii\omega t-\ii\bm{k}\cdot\bm{x}}\right),
\int\frac{\dd^{3}\bm{k}'}{(2\pi)^{3/2}}\frac{1}{\sqrt{2\omega'}}\left(\hat{a}_{\bm{k}'}^{\vphantom{\dagger}}e^{-\ii\omega' t'+\ii\bm{k}'\cdot\bm{x}'}+\hat{a}_{\bm{k}'}^{\dagger}e^{\ii\omega' t'-\ii\bm{k}'\cdot\bm{x}'}\right)\right],\\
&=\int\frac{\dd^{3}\bm{k}\dd^{3}\bm{k}'}{(2\pi)^{3}}\frac{\delta^{(3)}(\bm{k}-\bm{k}')}{\sqrt{4\omega\omega'}}\left(e^{-\ii(\omega t-\omega' t')+\ii(\bm{k}\cdot\bm{x}-\bm{k}'\cdot\bm{x}')}-e^{\ii(\omega t-\omega' t')-\ii(\bm{k}\cdot\bm{x}-\bm{k}'\cdot\bm{x}')}\right),\\
&=\int\frac{\dd^{3}\bm{k}}{(2\pi)^{3}}\frac{1}{2\omega}\left(e^{-\ii\omega(t-t')+\ii\bm{k}\cdot(\bm{x}-\bm{x}')}-e^{\ii\omega(t-t')-\ii\bm{k}\cdot(\bm{x}-\bm{x}')}\right).
\end{align}
In the massless case, we know that this commutator integrates to give delta functions along the lightcone, i.e. $\abs{\bm{x}-\bm{x}'}=\pm c(t-t')$. By inspection we can also see that the commutator above is on-shell, i.e. $\omega^{2}-\abs{\bm{k}}^{2}=0$. Applying the covariant UV cutoff $\Pi_{\Lambda}$ will therefore leave this commutator unchanged. This should not be surprising given that the field operators are on-shell already. 

In the main text above the trivial action of the covariant bandlimit on the commutator is said to be an issue regarding communication and micro-causality. 


Micro-causality, or microscopic causality, refers to the requirement that local field operators $\hat{\phi}(x)$ and $\hat{\pi}(y)$ must commute if $x$ and $y$ are space-like separated. A toy example of the importance of micro-causality involves two qubits A and B. Let $\{\hat{\mathcal{O}}_{\textsc{a,b}}^{i}\}$ be the set of Hermitian operators on A and B respectively such that $[\hat{\mathcal{O}}_{\textsc{a}}^{i},\hat{\mathcal{O}}_{\textsc{b}}^{j}]=0$, i.e. A and B are micro-causally space-like separated. Now assume that A performs some operation $\hat{U}_{\textsc{a}}$ on qubit A and B performs a measurement $\hat{\mathcal{P}}_{\textsc{b}}$ on qubit B (assume this is a POVM type measurement, i.e. B interacts with an ancilla and projectively measure the ancilla). Then, using a slight abuse of notation,
\begin{align}
\hat{U}_{\textsc{a}}\hat{\mathcal{P}}_{\textsc{b}}\ket{\Psi_{\textsc{a,b}}}=\hat{\mathcal{P}}_{\textsc{b}}\hat{U}_{\textsc{a}}\ket{\Psi_{\textsc{a,b}}},
\end{align}
i.e. since the set of Hermitian operators commute, their respective local operations commute. Consequently, it does not matter if B performs the measurement before A's operation or after A's operation. This tells us that A cannot communicate with B with the given set of operators and observables. Note that the toy model does not make any assumptions on the underlying spacetime structure of the model. 

The toy example highlights the importance of micro-causality in QFT, namely, micro-causality is a sufficient condition to ensure that two agents cannot communicate if they are space-like separated. 


The issue that arises in this manuscript is that the covariant UV cutoff does not modify the commutation relations, and therefore micro-causality appears to be preserved. However, as shown in \S\ref{sec3a}, acausal communication is possible in a covariantly cutoff QFT. From the toy model, we can see that the covariant UV cutoff does not break micro-causality, but rather it changes the set of operators that we assumed were local. Consider a non-cutoff theory where the set $\{\hat{\mathcal{O}}_{\textsc{a}}^{i}\}$ consists of local operators that act only on region A. Under the covariant UV cutoff, at least some of these operators $\hat{\mathcal{O}}_{\textsc{a}}^{i}$ will become nonlocal, i.e. act on regions beyond A. In the manuscript above we refer to these as the virtual particles, whose causal behavior can no longer be described using real particle commutation relations. 

\subsection{Field expectation values under covariant UV cutoffs}\label{sec_a4}
The introduction of a covariant UV cutoff causes confusion regarding energy carriers vs information carriers. The two-qubit communication protocol is used in this manuscript to show that covariant bandlimitation introduces acausal communication. Na\"{i}vely one would expect the expectation values of the field operators to share this acausal behavior. 

Consider a simple interaction Hamiltonian
\begin{align}
\hat{H}_{\textsc{i}}(t)&=\lambda \chi(t)\int\dd^{3}\bm{x}\,F(\bm{x})\hat{\phi}(t,\bm{x}),
\end{align}
where $\chi(t)$ and $F(\bm{x})$ are smooth compact functions. Note that this is not the UDW interaction Hamiltonian as this does not couple to a qubit. This Hamiltonian can then be used to perturbatively evolve the quantum field (initially in the vacuum) to
\begin{align}
\ket{\Psi(T)}&=\mathcal{T}\exp\left[-\ii\int\limits_{-\infty}^{T}\dd t\,\hat{H}_{\textsc{i}}(t)\right]\ket{0},\\
&=\left[\mathbb{I}-\ii\int\limits_{-\infty}^{T}\dd s\,\hat{H}_{\textsc{i}}(s)-
\int\limits_{-\infty}^{T}\dd s\int\limits_{-\infty}^{T}\dd s'\,\Theta(s-s')\hat{H}_{\textsc{i}}(s)\hat{H}_{\textsc{i}}(s')+\mathcal{O}(\lambda^{3})\right]\ket{0},
\end{align}
where we only time evolve up to the time $T$. If we now consider the expectation value of some local field operator ($\xi(y)$ some smooth compact function in spacetime):
\begin{align}
\bra{\Psi(T)}\int\dd^{4}y\,\xi(y)\hat{\phi}(y)\ket{\Psi(t)}&=
\bra{0}
\left[\mathbb{I}+\ii\int\limits_{-\infty}^{T}\dd s\,\hat{H}_{\textsc{i}}(s)-
\int\limits_{-\infty}^{T}\dd s\int\limits_{-\infty}^{T}\dd s'\,\Theta(s-s')\hat{H}_{\textsc{i}}(s')\hat{H}_{\textsc{i}}(s)\right]
\int\dd^{4}y\,\xi(y)\hat{\phi}(y)\nonumber\\
&\quad\times\left[\mathbb{I}-\ii\int\limits_{-\infty}^{T}\dd s\,\hat{H}_{\textsc{i}}(s)-
\int\limits_{-\infty}^{T}\dd s\int\limits_{-\infty}^{T}\dd s'\,\Theta(s-s')\hat{H}_{\textsc{i}}(s)\hat{H}_{\textsc{i}}(s')\right]\ket{0}.
\end{align}
In a non-cutoff theory, we expect this expectation value to be independent of $\lambda$ if there are no backward propagating time-like or light-like rays from the support of $\xi(y)$ that intersect with the support of $\chi(t)F(\bm{x})$, i.e. if the perturbation generated by the interaction Hamiltonian is space-like separated from the measurement device. 

In the covariantly bandlimited model, the two-qubit communication protocol suggests that this expectation value should be influenced by the acausal behavior of the model. However, an inspection of the equation above shows that there is no time-ordering between the measurement operator $\hat{\phi}(y)$ and the source terms in the Dyson series. This means that when evaluating the vacuum expectation value of the equation above there will not be any Feynman propagator or Green's function that has $y$ as a variable. Instead, there is an on-shell Wightman function with $y$ as a variable. Consequently, the introduction of the covariant bandlimit will not modify the causal structure of the equation above. 

For this reason, we say that these field expectation values are only sensitive to real particle excitations. However, from a Green's function perspective, this is a result of some incompleteness of the measurement theory, specifically our assumption that the field measurement algebra remained unchanged under the covariant UV cutoff.

\section{Two-qubit protocols}\label{sec_b0}
When considering covariant UV cutoffs, two-qubit protocols are the simplest that show observable changes in 2nd-order perturbation theory. 
\subsection{Two-qubit communication}\label{sec_b1}
The two-qubit communication protocol is an alternative perspective on the Fermi problem~\cite{Fermi_prob}, i.e. in the presence of atom A, how does the excitation probability of atom B change? In a causal theory, the answer should be zero if A and B are space-like separated. It is therefore a useful tool to evaluate if a field theory is causality-violating. 

In a causal theory, if Alice and Bob are space-like separated, i.e. the spacetime support of their interactions $\chi_{i}(t)F_{i}(\bm{x})$ have no time-like or light-like paths connecting them, an inspection of the reduced density matrix of either Alice or Bob should show no dependence on the interaction strength or initial state of the other qubit. Consider the post-interaction state \eqref{eqa12}. For simplicity we consider the initial state of the field to be the vacuum. Therefore, all the linear expectation values are zero. Then,
\begin{align}
\begin{split}
\hat{\rho}_{d,f}&=\left[\vphantom{\int\limits_{-\infty}^{\infty}}\right.
\hat{\rho}_{d,0}+\int\hspace{-2mm}\int\limits_{\mathbb{R}^{2}} \dd s\dd s' \left(
 \hat{\sigma}_{x}^{\textsc{a}}(s)\hat{\rho}_{d,0}\hat{\sigma}_{x}^{\textsc{a}}(s') \left<\vphantom{\phi}\right.\hspace{-0.6mm}\hat{\Phi}_{\textsc{a}}(s')\hat{\Phi}_{\textsc{a}}(s)\left.\hspace{-0.6mm}\vphantom{\phi}\right>
+\hat{\sigma}_{x}^{\textsc{a}}(s)\hat{\rho}_{d,0}\hat{\sigma}_{x}^{\textsc{b}}(s') \left<\vphantom{\phi}\right.\hspace{-0.6mm}\hat{\Phi}_{\textsc{a}}(s')\hat{\Phi}_{\textsc{b}}(s)\left.\hspace{-0.6mm}\vphantom{\phi}\right>\right.\\
&\quad\left.+\hat{\sigma}_{x}^{\textsc{b}}(s)\hat{\rho}_{d,0}\hat{\sigma}_{x}^{\textsc{a}}(s') \left<\vphantom{\phi}\right.\hspace{-0.6mm}\hat{\Phi}_{\textsc{b}}(s')\hat{\Phi}_{\textsc{a}}(s)\left.\hspace{-0.6mm}\vphantom{\phi}\right>
+\hat{\sigma}_{x}^{\textsc{b}}(s)\hat{\rho}_{d,0}\hat{\sigma}_{x}^{\textsc{b}}(s') \left<\vphantom{\phi}\right.\hspace{-0.6mm}\hat{\Phi}_{\textsc{b}}(s')\hat{\Phi}_{\textsc{b}}(s)\left.\hspace{-0.6mm}\vphantom{\phi}\right>
\right)\\
&\quad -\int\hspace{-2mm}\int\limits_{\mathbb{R}^{2}} \dd s\dd s'\,\Theta(s-s')\left(
\left[
 \hat{\sigma}_{x}^{\textsc{a}}(s)\hat{\sigma}_{x}^{\textsc{a}}(s')
\left<\vphantom{\phi}\right.\hspace{-0.6mm}\hat{\Phi}_{\textsc{a}}(s)\hat{\Phi}_{\textsc{a}}(s')\left.\hspace{-0.6mm}\vphantom{\phi}\right>
+\hat{\sigma}_{x}^{\textsc{a}}(s)\hat{\sigma}_{x}^{\textsc{b}}(s')
\left<\vphantom{\phi}\right.\hspace{-0.6mm}\hat{\Phi}_{\textsc{a}}(s)\hat{\Phi}_{\textsc{b}}(s')\left.\hspace{-0.6mm}\vphantom{\phi}\right>
+\hat{\sigma}_{x}^{\textsc{b}}(s)\hat{\sigma}_{x}^{\textsc{a}}(s')
\left<\vphantom{\phi}\right.\hspace{-0.6mm}\hat{\Phi}_{\textsc{b}}(s)\hat{\Phi}_{\textsc{a}}(s')\left.\hspace{-0.6mm}\vphantom{\phi}\right>\right.\right.\\
&\quad\left.+\hat{\sigma}_{x}^{\textsc{b}}(s)\hat{\sigma}_{x}^{\textsc{b}}(s')
\left<\vphantom{\phi}\right.\hspace{-0.6mm}\hat{\Phi}_{\textsc{b}}(s)\hat{\Phi}_{\textsc{b}}(s')\left.\hspace{-0.6mm}\vphantom{\phi}\right>\right]\hat{\rho}_{d,0}
+\hat{\rho}_{d,0}\left[
 \hat{\sigma}_{x}^{\textsc{a}}(s')\hat{\sigma}_{x}^{\textsc{a}}(s)
\left<\vphantom{\phi}\right.\hspace{-0.6mm}\hat{\Phi}_{\textsc{a}}(s')\hat{\Phi}_{\textsc{a}}(s)\left.\hspace{-0.6mm}\vphantom{\phi}\right>
+\hat{\sigma}_{x}^{\textsc{a}}(s')\hat{\sigma}_{x}^{\textsc{b}}(s)
\left<\vphantom{\phi}\right.\hspace{-0.6mm}\hat{\Phi}_{\textsc{a}}(s')\hat{\Phi}_{\textsc{b}}(s)\left.\hspace{-0.6mm}\vphantom{\phi}\right>\right.\\
&\quad\left.\left.+\hat{\sigma}_{x}^{\textsc{b}}(s')\hat{\sigma}_{x}^{\textsc{a}}(s)
\left<\vphantom{\phi}\right.\hspace{-0.6mm}\hat{\Phi}_{\textsc{b}}(s')\hat{\Phi}_{\textsc{a}}(s)\left.\hspace{-0.6mm}\vphantom{\phi}\right>
+\hat{\sigma}_{x}^{\textsc{b}}(s')\hat{\sigma}_{x}^{\textsc{b}}(s)
\left<\vphantom{\phi}\right.\hspace{-0.6mm}\hat{\Phi}_{\textsc{b}}(s')\hat{\Phi}_{\textsc{b}}(s)\left.\hspace{-0.6mm}\vphantom{\phi}\right>\right]\right)
\left.\vphantom{\int\limits_{-\infty}^{\infty}}\right].
\end{split}
\end{align}

We make the reasonable assumption that Alice and Bob are initially separable. Without loss of generality, we can take Alice to interact first and Bob later. As such we trace out Alice and consider the reduced density matrix of Bob:

\begin{align}
\begin{split}
\hat{\rho}_{\textsc{b},f}&=\left[\vphantom{\int\limits_{-\infty}^{\infty}}\right.
\hat{\rho}_{\textsc{b},0}+\int\hspace{-2mm}\int\limits_{\mathbb{R}^{2}} \dd s\dd s' \left(
 \left<\vphantom{\phi}\right.\hspace{-0.6mm}\hat{\sigma}_{x}^{\textsc{a}}(s') \hat{\sigma}_{x}^{\textsc{a}}(s)
\left.\hspace{-0.6mm}\vphantom{\phi}\right>\hat{\rho}_{\textsc{b},0}\left<\vphantom{\phi}\right.\hspace{-0.6mm}\hat{\Phi}_{\textsc{a}}(s')\hat{\Phi}_{\textsc{a}}(s)\left.\hspace{-0.6mm}\vphantom{\phi}\right>
+\left<\vphantom{\phi}\right.\hspace{-0.6mm}\hat{\sigma}_{x}^{\textsc{a}}(s)
\left.\hspace{-0.6mm}\vphantom{\phi}\right>
\hat{\rho}_{\textsc{b},0}\hat{\sigma}_{x}^{\textsc{b}}(s') \left<\vphantom{\phi}\right.\hspace{-0.6mm}\hat{\Phi}_{\textsc{a}}(s')\hat{\Phi}_{\textsc{b}}(s)\left.\hspace{-0.6mm}\vphantom{\phi}\right>\right.\\
&\quad\left.+\hat{\sigma}_{x}^{\textsc{b}}(s)\hat{\rho}_{\textsc{b},0}
\left<\vphantom{\phi}\right.\hspace{-0.6mm}\hat{\sigma}_{x}^{\textsc{a}}(s') 
\left.\hspace{-0.6mm}\vphantom{\phi}\right>
\left<\vphantom{\phi}\right.\hspace{-0.6mm}\hat{\Phi}_{\textsc{b}}(s')\hat{\Phi}_{\textsc{a}}(s)\left.\hspace{-0.6mm}\vphantom{\phi}\right>
+\hat{\sigma}_{x}^{\textsc{b}}(s)\hat{\rho}_{\textsc{b},0}\hat{\sigma}_{x}^{\textsc{b}}(s') \left<\vphantom{\phi}\right.\hspace{-0.6mm}\hat{\Phi}_{\textsc{b}}(s')\hat{\Phi}_{\textsc{b}}(s)\left.\hspace{-0.6mm}\vphantom{\phi}\right>
\right)\\
&\quad -\int\hspace{-2mm}\int\limits_{\mathbb{R}^{2}} \dd s\dd s'\,\Theta(s-s')\left(
\left[
 \left<\vphantom{\phi}\right.\hspace{-0.6mm}\hat{\sigma}_{x}^{\textsc{a}}(s)\hat{\sigma}_{x}^{\textsc{a}}(s')\left.\hspace{-0.6mm}\vphantom{\phi}\right>
\left<\vphantom{\phi}\right.\hspace{-0.6mm}\hat{\Phi}_{\textsc{a}}(s)\hat{\Phi}_{\textsc{a}}(s')\left.\hspace{-0.6mm}\vphantom{\phi}\right>
+\left<\vphantom{\phi}\right.\hspace{-0.6mm}\hat{\sigma}_{x}^{\textsc{a}}(s)\left.\hspace{-0.6mm}\vphantom{\phi}\right>\hat{\sigma}_{x}^{\textsc{b}}(s')
\left<\vphantom{\phi}\right.\hspace{-0.6mm}\hat{\Phi}_{\textsc{a}}(s)\hat{\Phi}_{\textsc{b}}(s')\left.\hspace{-0.6mm}\vphantom{\phi}\right>
+\hat{\sigma}_{x}^{\textsc{b}}(s)\left<\vphantom{\phi}\right.\hspace{-0.6mm}\hat{\sigma}_{x}^{\textsc{a}}(s')\left.\hspace{-0.6mm}\vphantom{\phi}\right>
\left<\vphantom{\phi}\right.\hspace{-0.6mm}\hat{\Phi}_{\textsc{b}}(s)\hat{\Phi}_{\textsc{a}}(s')\left.\hspace{-0.6mm}\vphantom{\phi}\right>\right.\right.\\
&\quad\left.+\hat{\sigma}_{x}^{\textsc{b}}(s)\hat{\sigma}_{x}^{\textsc{b}}(s')
\left<\vphantom{\phi}\right.\hspace{-0.6mm}\hat{\Phi}_{\textsc{b}}(s)\hat{\Phi}_{\textsc{b}}(s')\left.\hspace{-0.6mm}\vphantom{\phi}\right>\right]\hat{\rho}_{d,0}
+\hat{\rho}_{d,0}\left[
 \left<\vphantom{\phi}\right.\hspace{-0.6mm}\hat{\sigma}_{x}^{\textsc{a}}(s')\hat{\sigma}_{x}^{\textsc{a}}(s)\left.\hspace{-0.6mm}\vphantom{\phi}\right>
\left<\vphantom{\phi}\right.\hspace{-0.6mm}\hat{\Phi}_{\textsc{a}}(s')\hat{\Phi}_{\textsc{a}}(s)\left.\hspace{-0.6mm}\vphantom{\phi}\right>
+\left<\vphantom{\phi}\right.\hspace{-0.6mm}\hat{\sigma}_{x}^{\textsc{a}}(s')\left.\hspace{-0.6mm}\vphantom{\phi}\right>\hat{\sigma}_{x}^{\textsc{b}}(s)
\left<\vphantom{\phi}\right.\hspace{-0.6mm}\hat{\Phi}_{\textsc{a}}(s')\hat{\Phi}_{\textsc{b}}(s)\left.\hspace{-0.6mm}\vphantom{\phi}\right>\right.\\
&\quad\left.\left.+\hat{\sigma}_{x}^{\textsc{b}}(s')\left<\vphantom{\phi}\right.\hspace{-0.6mm}\hat{\sigma}_{x}^{\textsc{a}}(s)\left.\hspace{-0.6mm}\vphantom{\phi}\right>
\left<\vphantom{\phi}\right.\hspace{-0.6mm}\hat{\Phi}_{\textsc{b}}(s')\hat{\Phi}_{\textsc{a}}(s)\left.\hspace{-0.6mm}\vphantom{\phi}\right>
+\hat{\sigma}_{x}^{\textsc{b}}(s')\hat{\sigma}_{x}^{\textsc{b}}(s)
\left<\vphantom{\phi}\right.\hspace{-0.6mm}\hat{\Phi}_{\textsc{b}}(s')\hat{\Phi}_{\textsc{b}}(s)\left.\hspace{-0.6mm}\vphantom{\phi}\right>\right]\right)
\left.\vphantom{\int\limits_{-\infty}^{\infty}}\right].
\end{split}
\end{align}
By considering the Heaviside step function $\Theta(t)+\Theta(-t)=1$ we can see that the $\lambda_{\textsc{a}}^{2}$ terms cancel out. After this cancellation we note that the remaining term are $\hat{\rho}_{\textsc{b},0}$, dependent on $\lambda_{\textsc{b}}^{2}$ or dependent on $\lambda_{\textsc{a}}\lambda_{\textsc{b}}$. Since we are interested in the effects of Alice on Bob's qubit we only consider the terms dependent on $\lambda_{\textsc{a}}\lambda_{\textsc{b}}$. We therefore find that
\begin{align}
\begin{split}
\hat{\rho}_{\textsc{b},f}&=\left[\vphantom{\int\limits_{-\infty}^{\infty}}\right.
\hat{\rho}_{\textsc{b},0}+\int\hspace{-2mm}\int\limits_{\mathbb{R}^{2}} \dd s\dd s' \left(
\left<\vphantom{\phi}\right.\hspace{-0.6mm}\hat{\sigma}_{x}^{\textsc{a}}(s)
\left.\hspace{-0.6mm}\vphantom{\phi}\right>
\hat{\rho}_{\textsc{b},0}\hat{\sigma}_{x}^{\textsc{b}}(s') \left<\vphantom{\phi}\right.\hspace{-0.6mm}\hat{\Phi}_{\textsc{a}}(s')\hat{\Phi}_{\textsc{b}}(s)\left.\hspace{-0.6mm}\vphantom{\phi}\right>+\hat{\sigma}_{x}^{\textsc{b}}(s)\hat{\rho}_{\textsc{b},0}
\left<\vphantom{\phi}\right.\hspace{-0.6mm}\hat{\sigma}_{x}^{\textsc{a}}(s') 
\left.\hspace{-0.6mm}\vphantom{\phi}\right>
\left<\vphantom{\phi}\right.\hspace{-0.6mm}\hat{\Phi}_{\textsc{b}}(s')\hat{\Phi}_{\textsc{a}}(s)\left.\hspace{-0.6mm}\vphantom{\phi}\right>
\right)\\
&\quad -\int\hspace{-2mm}\int\limits_{\mathbb{R}^{2}} \dd s\dd s'\,\Theta(s-s')\left(
\left[
\left<\vphantom{\phi}\right.\hspace{-0.6mm}\hat{\sigma}_{x}^{\textsc{a}}(s)\left.\hspace{-0.6mm}\vphantom{\phi}\right>\hat{\sigma}_{x}^{\textsc{b}}(s')
\left<\vphantom{\phi}\right.\hspace{-0.6mm}\hat{\Phi}_{\textsc{a}}(s')\hat{\Phi}_{\textsc{b}}(s)\left.\hspace{-0.6mm}\vphantom{\phi}\right>
+\hat{\sigma}_{x}^{\textsc{b}}(s)\left<\vphantom{\phi}\right.\hspace{-0.6mm}\hat{\sigma}_{x}^{\textsc{a}}(s')\left.\hspace{-0.6mm}\vphantom{\phi}\right>
\left<\vphantom{\phi}\right.\hspace{-0.6mm}\hat{\Phi}_{\textsc{b}}(s')\hat{\Phi}_{\textsc{a}}(s)\left.\hspace{-0.6mm}\vphantom{\phi}\right>\right]\right.\hat{\rho}_{d,0}\\
&\quad\left.+\hat{\rho}_{d,0}\left[
 \left<\vphantom{\phi}\right.\hspace{-0.6mm}\hat{\sigma}_{x}^{\textsc{a}}(s')\left.\hspace{-0.6mm}\vphantom{\phi}\right>\hat{\sigma}_{x}^{\textsc{b}}(s)
\left<\vphantom{\phi}\right.\hspace{-0.6mm}\hat{\Phi}_{\textsc{a}}(s')\hat{\Phi}_{\textsc{b}}(s)\left.\hspace{-0.6mm}\vphantom{\phi}\right>+\hat{\sigma}_{x}^{\textsc{b}}(s')\left<\vphantom{\phi}\right.\hspace{-0.6mm}\hat{\sigma}_{x}^{\textsc{a}}(s)\left.\hspace{-0.6mm}\vphantom{\phi}\right>
\left<\vphantom{\phi}\right.\hspace{-0.6mm}\hat{\Phi}_{\textsc{b}}(s')\hat{\Phi}_{\textsc{a}}(s)\left.\hspace{-0.6mm}\vphantom{\phi}\right>
\right]\right)
\left.\vphantom{\int\limits_{-\infty}^{\infty}}\right]+\mathcal{O}(\lambda_{\textsc{b}}^{2}).
\end{split}
\end{align}
In non-bandlimited theories, the next step would be to observe that Alice interacted before Bob and therefore \\
$\Theta(s-s')\chi_{\textsc{a}}(s)\chi_{\textsc{b}}(s')\rightarrow 0$. However, our covariant bandlimitation means this is no longer true. 

A further simplification is to consider Alice and Bob's initial state as pure and separable $\ket{\psi_{0}}=\ket{\psi_{\textsc{a}}}\otimes\ket{\psi_{\textsc{b}}}\otimes\ket{0}$ and then also perform a projective measurement on Bob $\bra{\psi}\hat{\rho}_{\textsc{b}}\ket{\psi}$. By introducing these simplifications one can derive \eqref{eq19} from the equations directly above. The crude form of $\mathcal{I}_{\Lambda}$ is
\begin{align}
\mathcal{I}_{\Lambda}(r,t)&=\frac{2}{\pi}\int\limits_{0}^{\infty}\dd\omega\,\sin(\omega r)\left\{
\frac{\text{P.V.}}{2\pi}\left[\,\int\limits_{\omega-\sqrt{\omega^{2}+\Lambda^{2}}}^{\rho^{-}}+\int\limits_{\rho^{+}}^{\omega+\sqrt{\omega^{2}+\Lambda^{2}}}\,\right]\frac{2}{\nu}\cos[(\omega-\nu)t]\,\dd\nu\right\},
\end{align}
where P.V. denotes principal part of the integral, $\rho^{\pm}=\omega\pm\sqrt{\omega^{2}-\Lambda^{2}}$ when $\omega>\Lambda$ and $\rho^{+}=\rho^{-}$ if $\omega\leq \Lambda$. $\mathcal{I}_{\Lambda}$ can be thought of as the principal value part of an integral, whose residues yield the Dirac deltas in \eqref{eq19}. As $\Lambda\rightarrow \infty$ this function goes to zero.

\subsection{Entanglement Harvesting}\label{sec_b2}
The two-qubit protocol described above is particularly useful for determining the causal structure of a field theory, whether it is classical or quantum. However, it is too simple a protocol to describe any quantum properties of the field. This is where entanglement harvesting comes in.

Entanglement harvesting is a two-qubit protocol, commonly used in RQI and capable of evaluating nonlocal quantum properties of a quantum field, such as entanglement or mutual information. Many works and useful reviews are available~\cite{PhysRevD.94.064074,PhysRevD.96.065008,PhysRevD.98.085007,PhysRevD.96.085012,PhysRevD.79.044027,robbins2020entanglement,HENDERSON2020135732,hendersonEntanglingDetectorsAntide2019,Martin-Martinez_2014}. Here we will review the basics of entanglement harvesting for completeness. 

Entanglement harvesting requires two UDW detectors~\cite{UDW_1,UDW_2} (generally qubits) that are usually space-like separated, i.e. their spacetime interaction regions (given by $\chi(t)F(\bm{x})$) are space-like separated. The quantum field is initially in the vacuum state (although other states have been studied~\cite{PhysRevD.98.085007}) and the detectors are initially in their respective ground states. Following the interaction of the detectors with the field, the field is traced out and the reduced density matrix of the two detectors is analyzed. Recall the interaction Hamiltonian for the UDW detector where $\lambda_{i}$ is the interaction strength, $\chi_{i}(t)$ describes when the detectors are on (switching) and $F_{i}(\bm{x})$ describes the size of the detectors (smearing):

\begin{align}
\hat{H}_{\textsc{i}}&=\sum_{i\in\{\textsc{a,b}\}}\lambda_{i}\chi_{i}(t)\hat{\sigma}^{i}_{x}(t)\int\dd^{3}\bm{x}\,F_{i}(\bm{x})\hat{\phi}(t,\bm{x}).
\end{align}

Consider \eqref{eqa12} under these requirements and recall that $\hat{\sigma}_{x}^{i}(t)=\hat{\sigma}^{+,i}e^{\ii\Omega_{i} t}+\hat{\sigma}^{-,i}e^{-\ii\Omega_{i} t}$, where $\Omega_{i}$ is the energy gap of qubit (detector) $i$ and $\hat{\sigma}^{\pm}$ are the raising and lowering operators of the qubit $i$:

\begin{align}
\begin{split}
\hat{\rho}_{d,f}&=\ketbra{0_{\textsc{a}}0_{\textsc{b}}}{0_{\textsc{a}}0_{\textsc{b}}}
+\int\dd s\dd s'\dd^{3}\bm{x}\dd^{3}\bm{x}'\,
\left(
\lambda_{\textsc{a}}^{2}
\chi_{\textsc{a}}(s)\chi_{\textsc{a}}(s')
F_{\textsc{a}}(\bm{x})F_{\textsc{a}}(\bm{x}')
e^{\ii\Omega_{\textsc{a}}(s-s')}\ketbra{1_{\textsc{a}}0_{\textsc{b}}}{1_{\textsc{a}}0_{\textsc{b}}}\right.\\
&\quad+\lambda_{\textsc{a}}\lambda_{\textsc{b}}
\chi_{\textsc{a}}(s)\chi_{\textsc{b}}(s')
F_{\textsc{a}}(\bm{x})F_{\textsc{b}}(\bm{x}')
e^{\ii(\Omega_{\textsc{a}}s-\Omega_{\textsc{b}}s')}\ketbra{1_{\textsc{a}}0_{\textsc{b}}}{0_{\textsc{a}}1_{\textsc{b}}}
+\lambda_{\textsc{b}}\lambda_{\textsc{a}}
\chi_{\textsc{b}}(s)\chi_{\textsc{a}}(s')
F_{\textsc{b}}(\bm{x})F_{\textsc{a}}(\bm{x}')
e^{\ii(\Omega_{\textsc{b}}s-\Omega_{\textsc{a}}s')}\ketbra{0_{\textsc{a}}1_{\textsc{b}}}{1_{\textsc{a}}0_{\textsc{b}}}\\
&\quad\left.+
\lambda_{\textsc{b}}^{2}
\chi_{\textsc{b}}(s)\chi_{\textsc{b}}(s')
F_{\textsc{b}}(\bm{x})F_{\textsc{b}}(\bm{x}')
e^{\ii\Omega_{\textsc{b}}(s-s')}\ketbra{0_{\textsc{a}}1_{\textsc{b}}}{0_{\textsc{a}}1_{\textsc{b}}}
\right)
\left<\vphantom{\phi}\right.\hspace{-0.6mm}
\hat{\phi}(s',\bm{x}')\hat{\phi}(s,\bm{x})
\left.\hspace{-0.6mm}\vphantom{\phi}\right>\\
&\quad-\int\dd s\dd s'\dd^{3}\bm{x}\dd^{3}\bm{x}'\,\Theta(s-s')
\left(
\lambda_{\textsc{a}}^{2}
\chi_{\textsc{a}}(s)\chi_{\textsc{a}}(s')
F_{\textsc{a}}(\bm{x})F_{\textsc{a}}(\bm{x}')
e^{-\ii\Omega_{\textsc{a}}(s-s')}\ketbra{0_{\textsc{a}}0_{\textsc{b}}}{0_{\textsc{a}}0_{\textsc{b}}}\right.\\
&\quad+\lambda_{\textsc{a}}\lambda_{\textsc{b}}
\chi_{\textsc{a}}(s)\chi_{\textsc{b}}(s')
F_{\textsc{a}}(\bm{x})F_{\textsc{b}}(\bm{x}')
e^{\ii(\Omega_{\textsc{a}}s+\Omega_{\textsc{b}}s')}\ketbra{1_{\textsc{a}}1_{\textsc{b}}}{0_{\textsc{a}}0_{\textsc{b}}}
+\lambda_{\textsc{b}}\lambda_{\textsc{a}}
\chi_{\textsc{b}}(s)\chi_{\textsc{a}}(s')
F_{\textsc{b}}(\bm{x})F_{\textsc{a}}(\bm{x}')
e^{\ii(\Omega_{\textsc{b}}s+\Omega_{\textsc{a}}s')}\ketbra{1_{\textsc{a}}1_{\textsc{b}}}{0_{\textsc{a}}0_{\textsc{b}}}\\
&\quad\left.+\lambda_{\textsc{b}}^{2}
\chi_{\textsc{b}}(s)\chi_{\textsc{b}}(s')
F_{\textsc{b}}(\bm{x})F_{\textsc{b}}(\bm{x}')
e^{-\ii\Omega_{\textsc{b}}(s-s')}\ketbra{0_{\textsc{a}}0_{\textsc{b}}}{0_{\textsc{a}}0_{\textsc{b}}}
\right)
\left<\vphantom{\phi}\right.\hspace{-0.6mm}
\hat{\phi}(s,\bm{x})\hat{\phi}(s',\bm{x}')
\left.\hspace{-0.6mm}\vphantom{\phi}\right>\\
&\quad-\int\dd s\dd s'\dd^{3}\bm{x}\dd^{3}\bm{x}'\,\Theta(s-s')
\left(
\lambda_{\textsc{a}}^{2}
\chi_{\textsc{a}}(s')\chi_{\textsc{a}}(s)
F_{\textsc{a}}(\bm{x}')F_{\textsc{a}}(\bm{x})
e^{\ii\Omega_{\textsc{a}}(s-s')}\ketbra{0_{\textsc{a}}0_{\textsc{b}}}{0_{\textsc{a}}0_{\textsc{b}}}\right.\\
&\quad+\lambda_{\textsc{a}}\lambda_{\textsc{b}}
\chi_{\textsc{a}}(s')\chi_{\textsc{b}}(s)
F_{\textsc{a}}(\bm{x}')F_{\textsc{b}}(\bm{x})
e^{-\ii(\Omega_{\textsc{a}}s'+\Omega_{\textsc{b}}s)}\ketbra{0_{\textsc{a}}0_{\textsc{b}}}{1_{\textsc{a}}1_{\textsc{b}}}
+\lambda_{\textsc{b}}\lambda_{\textsc{a}}
\chi_{\textsc{b}}(s')\chi_{\textsc{a}}(s)
F_{\textsc{b}}(\bm{x}')F_{\textsc{a}}(\bm{x})
e^{-\ii(\Omega_{\textsc{a}}s+\Omega_{\textsc{b}}s')}\ketbra{0_{\textsc{a}}0_{\textsc{b}}}{1_{\textsc{a}}1_{\textsc{b}}}\\
&\quad\left.+\lambda_{\textsc{b}}^{2}
\chi_{\textsc{b}}(s')\chi_{\textsc{b}}(s)
F_{\textsc{b}}(\bm{x}')F_{\textsc{b}}(\bm{x})
e^{\ii\Omega_{\textsc{b}}(s-s')}\ketbra{0_{\textsc{a}}0_{\textsc{b}}}{0_{\textsc{a}}0_{\textsc{b}}}\right)
\left<\vphantom{\phi}\right.\hspace{-0.6mm}
\hat{\phi}(s',\bm{x}')\hat{\phi}(s,\bm{x})
\left.\hspace{-0.6mm}\vphantom{\phi}\right>,
\end{split}\\
\begin{split}
&=\ketbra{0_{\textsc{a}}0_{\textsc{b}}}{0_{\textsc{a}}0_{\textsc{b}}}\\
&\quad-\sum_{i\in\{\textsc{a,b}\}}\int \dd s \dd s'\dd^{3}\bm{x}\dd^{3}\bm{x}'\, (\Theta(s-s')+\Theta(s'-s))\lambda_{i}^{2}\chi_{i}(s)\chi_{i}(s')F_{i}(\bm{x})F_{i}(\bm{x}')e^{\ii\Omega_{i}(s-s')}
\left<\vphantom{\phi}\right.\hspace{-0.6mm}
\hat{\phi}(s',\bm{x}')\hat{\phi}(s,\bm{x})
\left.\hspace{-0.6mm}\vphantom{\phi}\right>
\ketbra{0_{\textsc{a}}0_{\textsc{b}}}{0_{\textsc{a}}0_{\textsc{b}}}\\
&\quad+\sum_{i,j\in\{\textsc{a,b}\}}\int \dd s \dd s'\dd^{3}\bm{x}\dd^{3}\bm{x}'\,
\lambda_{i}\lambda_{j}\chi_{i}(s)\chi_{j}(s')F_{i}(\bm{x})F_{j}(\bm{x}')e^{\ii(\Omega_{i}s-\Omega_{j}s')}
\left<\vphantom{\phi}\right.\hspace{-0.6mm}
\hat{\phi}(s',\bm{x}')\hat{\phi}(s,\bm{x})
\left.\hspace{-0.6mm}\vphantom{\phi}\right>
\, \hat{\sigma}^{+,i}
\ketbra{0_{\textsc{a}}0_{\textsc{b}}}{0_{\textsc{a}}0_{\textsc{b}}}
\hat{\sigma}^{-,j}\\
&\quad-\int\dd s\dd s'\dd^{3}\bm{x}\dd^{3}\bm{x}'\,\Theta(s-s')\lambda_{\textsc{a}}\lambda_{\textsc{b}}F_{\textsc{a}}(\bm{x})F_{\textsc{b}}(\bm{x}')
\left(
\chi_{\textsc{a}}(s)\chi_{\textsc{b}}(s')e^{\ii(\Omega_{\textsc{a}}s+\Omega_{\textsc{b}}s')}
\left<\vphantom{\phi}\right.\hspace{-0.6mm}
\hat{\phi}(s,\bm{x})\hat{\phi}(s',\bm{x}')
\left.\hspace{-0.6mm}\vphantom{\phi}\right>\right.\\
&\quad\left.+\chi_{\textsc{b}}(s)\chi_{\textsc{a}}(s')e^{\ii(\Omega_{\textsc{b}}s+\Omega_{\textsc{a}}s')}
\left<\vphantom{\phi}\right.\hspace{-0.6mm}
\hat{\phi}(s,\bm{x}')\hat{\phi}(s',\bm{x})
\left.\hspace{-0.6mm}\vphantom{\phi}\right>
\right)
\ketbra{1_{\textsc{a}}1_{\textsc{b}}}{0_{\textsc{a}}0_{\textsc{b}}}\\
&\quad-\int\dd s\dd s'\dd^{3}\bm{x}\dd^{3}\bm{x}'\,\Theta(s-s')\lambda_{\textsc{a}}\lambda_{\textsc{b}}F_{\textsc{a}}(\bm{x})F_{\textsc{b}}(\bm{x}')
\left(
\chi_{\textsc{a}}(s)\chi_{\textsc{b}}(s')e^{-\ii(\Omega_{\textsc{a}}s+\Omega_{\textsc{b}}s')}
\left<\vphantom{\phi}\right.\hspace{-0.6mm}
\hat{\phi}(s',\bm{x}')\hat{\phi}(s,\bm{x})
\left.\hspace{-0.6mm}\vphantom{\phi}\right>\right.\\
&\quad\left.+\chi_{\textsc{a}}(s')\chi_{\textsc{b}}(s)e^{-\ii(\Omega_{\textsc{a}}s'+\Omega_{\textsc{b}}s)}
\left<\vphantom{\phi}\right.\hspace{-0.6mm}
\hat{\phi}(s',\bm{x})\hat{\phi}(s,\bm{x}')
\left.\hspace{-0.6mm}\vphantom{\phi}\right>
\right)
\ketbra{0_{\textsc{a}}0_{\textsc{b}}}{1_{\textsc{a}}1_{\textsc{b}}},
\end{split}\\
\begin{split}
&=(1-\mathcal{L}_{\textsc{aa}}-\mathcal{L}_{\textsc{bb}})
\ketbra{0_{\textsc{a}}0_{\textsc{b}}}{0_{\textsc{a}}0_{\textsc{b}}}
+\sum_{i,j\in\{\textsc{a,b}\}}\mathcal{L}_{ij}\hat{\sigma}^{+,i}
\ketbra{0_{\textsc{a}}0_{\textsc{b}}}{0_{\textsc{a}}0_{\textsc{b}}}
\hat{\sigma}^{-,j}
+\mathcal{M}\ketbra{1_{\textsc{a}}1_{\textsc{b}}}{0_{\textsc{a}}0_{\textsc{b}}}
+\mathcal{M}^{*}
\ketbra{0_{\textsc{a}}0_{\textsc{b}}}{1_{\textsc{a}}1_{\textsc{b}}}.
\end{split}
\end{align}
That is
\begin{align}
\hat{\rho}_{\textsc{ab}}&=
\begin{pmatrix}
1-\mathcal{L}_{\textsc{aa}}-\mathcal{L}_{\textsc{bb}} & 0 & 0 & \mathcal{M}^{*}\\
0 & \mathcal{L}_{\textsc{aa}} & \mathcal{L}_{\textsc{ab}} & 0 \\
0 & \mathcal{L}_{\textsc{ba}} & \mathcal{L}_{\textsc{bb}} & 0 \\
\mathcal{M} & 0 & 0 & 0
\end{pmatrix}
+\mathcal{O}(\lambda^{4}),
\end{align}
where
\begin{align}
\mathcal{L}_{ij}&=\lambda_{i}\lambda_{j}\int\dd s\dd s'\dd^{3}\bm{x}\dd^{3}\bm{x}'\,e^{\ii\Omega_{i}s-\ii\Omega_{j}s'}\chi_{i}(s)\chi_{j}(s')\nonumber\\
&\quad\times F_{i}(\bm{x})F_{j}(\bm{x}')
\left<\vphantom{\phi}\right.\hspace{-0.6mm}\hat{\phi}(s',\bm{x}')\hat{\phi}(s,\bm{x})\left.\hspace{-0.6mm}\vphantom{\phi}\right>,\\
\mathcal{M}&=-\lambda_{\textsc{a}}\lambda_{\textsc{b}}\int\dd s\dd s'\dd^{3}\bm{x}\dd^{3}\bm{x}'\,
\Theta(s-s') F_{\textsc{a}}(\bm{x})F_{\textsc{b}}(\bm{x}')\nonumber\\
&\quad\times\bigg\{
e^{\ii\Omega_{\textsc{a}}s+\ii\Omega_{\textsc{b}}s'} \chi_{\textsc{a}}(s)\chi_{\textsc{b}}(s')
\left<\vphantom{\phi}\right.\hspace{-0.6mm}\hat{\phi}(s,\bm{x})\hat{\phi}(s',\bm{x}')\left.\hspace{-0.6mm}\vphantom{\phi}\right>\nonumber\\
&\quad+e^{\ii\Omega_{\textsc{b}}s+\ii\Omega_{\textsc{a}}s'} \chi_{\textsc{b}}(s)\chi_{\textsc{a}}(s')
\left<\vphantom{\phi}\right.\hspace{-0.6mm}\hat{\phi}(s,\bm{x}')\hat{\phi}(s',\bm{x})\left.\hspace{-0.6mm}\vphantom{\phi}\right>
\bigg\}.
\end{align}

Finally, from the reduced density matrix we can evaluate the entanglement negativity $\mathcal{N}=\operatorname{max}(0,\mathcal{N}^{(2)})$, where~\cite{PhysRevA.71.042104}
\begin{align}
\mathcal{N}^{(2)}&=-\frac{1}{2}\left(
\mathcal{L}_{\textsc{aa}}+\mathcal{L}_{\textsc{bb}}
-\sqrt{\left(\mathcal{L}_{\textsc{aa}}-\mathcal{L}_{\textsc{bb}}\right)^{2}+4\abs{\mathcal{M}}^{2}}\right).
\end{align}









\twocolumngrid
\bibliography{ref}

\end{document}